%% file: main.tex
\documentclass[12pt]{article}
\pdfoutput=1
\usepackage[margin=3cm]{geometry}
\usepackage[utf8]{inputenc}
\usepackage{mathptmx}

\usepackage{xcolor}
\definecolor{tealblue}{rgb}{0.21, 0.46, 0.53}
\usepackage{hyperref}
\hypersetup{colorlinks=true,allcolors = tealblue,linktocpage=true}

\usepackage{amsmath}
\usepackage{amssymb}
\usepackage{authblk} 
\usepackage{physics}
\usepackage{graphicx} 

\usepackage{pgf} 

\usepackage[
        backend=bibtex,
        sorting=none,
        style=phys,
        eprint=true,
        doi=false,
        biblabel=brackets
        ]{biblatex}

\bibliography{brane_dynamics.bib}

\newcommand{\ds}{\ensuremath{\operatorname{dS}}}
\newcommand{\cft}{\ensuremath{\operatorname{CFT}}}

\newcommand{\commie}[1]{}

\newcommand{\ads}{\mathrm{AdS}}
\newcommand{\ess}{\mathbb{S}}
\newcommand{\Vol}{\mathrm{Vol}}
\newcommand{\adsts}{{\ads \times \ess}}

\numberwithin{equation}{section}

\newenvironment{eqaed}
    {\begin{equation}
    \begin{aligned}
    }
    { 
    \end{aligned}
    \end{equation}
    }

\begin{document}

\title{\bf Supersymmetry breaking, brane dynamics and Swampland conjectures}

\author{Ivano Basile\thanks{ivano.basile@umons.ac.be}}
\affil{\emph{Service de Physique de l'Univers, Champs et Gravitation\\ \emph{Universit\'{e} de Mons}}\\\emph{Place du Parc 20, 7000 Mons, Belgium}}

\maketitle

\begin{abstract}
     
     We investigate interactions between branes of various dimensions, both charged and uncharged, in three non-supersymmetric string models. These include the $USp(32)$ and $U(32)$ orientifold projections of the type IIB and type $0$B strings, as well as the $SO(16) \times SO(16)$ projection of the exceptional heterotic string. The resulting ten-dimensional spectra are free of tachyons, and the combinations of branes that they contain give rise to rich and varied dynamics. We compute static potentials for parallel stacks of branes in three complementary regimes: the probe regime, in which one of the two stacks is parametrically heavier than the other, the string-amplitude regime, in which both stacks are light, and the holographic regime. Whenever comparisons are possible, we find qualitative agreement despite the absence of supersymmetry. For charged branes, our analysis reveals that the Weak Gravity Conjecture is satisfied in a novel way via a renormalization of the effective charge-to-tension ratio.
    
\end{abstract}

\tableofcontents

\section{Introduction}\label{sec:introduction}
    
    Achieving a deeper understanding of string theory, and quantum gravity in general, as well as building stronger ties with phenomenology, requires addressing the problem of supersymmetry breaking. Despite remarkable progress has been accomplished in a variety of directions, the lack of an overarching guiding principle together with numerous issues in low-energy effective field theory (EFT) constructions points to high-energy supersymmetry breaking as the natural environment to seek instructive lessons, both from a theoretical and a phenomenological perspective. Indeed, it is possible to concoct perturbative models in which supersymmetry is either absent or broken at the string scale. Within the scope of this paper, the former setting comprise the $SO(16) \times SO(16)$ heterotic model of~\cite{AlvarezGaume:1986jb, Dixon:1986iz} and the $U(32)$ ``type $0'$B'' model of~\cite{Sagnotti:1995ga, Sagnotti:1996qj}, while the latter setting is embodied by the $USp(32)$ model of~\cite{Sugimoto:1999tx}, which exhibits the peculiar phenomenon of ``brane supersymmetry breaking'' (BSB)~\cite{Antoniadis:1999xk, Angelantonj:1999jh, Aldazabal:1999jr, Angelantonj:1999ms}\footnote{See also~\cite{Coudarchet:2021qwc} for a recent investigation of models featuring a novel type of BSB.}. The notion that high-energy supersymmetry breaking is in some sense natural also resonates with recent considerations stemming from the Swampland program~\cite{Cribiori:2021gbf, Castellano:2021yye}, and in particular the central role played by the gravitino mass is suggestively reminiscent of BSB, as discussed in~\cite{Coudarchet:2021qwc}.
    
    At any rate, in order to elucidate deeper features of supersymmetry breaking it appears paramount to go beyond the EFT regime. Specifically, one can consider the dynamics of branes, which manifest themselves in various guises in different regimes of string theory, ranging from soliton solutions of space-time field equations to conformal field theory (CFT) boundary states to world-volume gauge theories. They have been proven an invaluable tool to investigate any setting, supersymmetric or not, and in this paper we would like to pursue this approach to shed some light on the subtle, and so far elusive, physics of high-energy supersymmetry breaking. Concretely, we shall compute static interaction potentials between stacks of parallel branes of various charges and dimensions, comparing the results obtained in different regimes. In particular, the interactions between extremal branes reveal a peculiar, novel mechanism giving rise to a repulsive force~\cite{Antonelli:2019nar}, which we shall discuss in detail in the following, further grounding the connections between top-down settings and the Swampland program that were developed in~\cite{Basile:2020mpt}.
    
    The contents of this paper are organized as follows. In Section~\ref{sec:string_models} we provide an overview of the non-supersymmetric string models at stake, describing their brane content in Section~\ref{sec:brane_scan} and low-energy EFT description in Section~\ref{sec:low-energy}. In Section~\ref{sec:geometries} we discuss the gravitational back-reaction of branes of various dimensions, namely uncharged $\text{D}8$-branes and charged $\text{D}1$, $\text{D}3$ and $\text{NS}5$-branes. The latter feature near-horizon Anti-de Sitter ($\ads$) throats, characteristic of extremal black objects, with the exception of $\text{D}3$-branes in the type $0'$B model of~\cite{Sagnotti:1995ga, Sagnotti:1996qj} which feature quasi-$\ads$ geometries that converge non-uniformly to $\ads_5 \times \ess^5$ for a large number of branes~\cite{Angelantonj:1999qg, Angelantonj:2000kh}. In Section~\ref{sec:probes} we present in detail the computation of interaction potentials for parallel brane stacks of various dimensions and charges. Our analysis spans three complementary regimes, depending on the numbers $N_1 \, , \, N_2$ of branes in the two stacks. In particular, for $N_1 \ll N_2$ the branes in the first stack effectively probe the space-time geometry sourced by the second stack, while for $N_1 \, , \, N_2 = \mathcal{O}(1)$ the interaction potential is calculable via perturbative string amplitudes, at least in principle. Whenever both stacks have the same dimensions and charges the resulting expressions are highly involved, but in all other cases the leading contribution for large brane separations can be extracted from the annulus amplitude. Finally, we study the world-volume gauge theory of $\text{D}1$-branes in the $USp(32)$ model of~\cite{Sugimoto:1999tx}, where the bosonic and fermionic contributions to the one-loop effective potential do not cancel, leading to a non-trivial result along the lines of~\cite{Zarembo:1999hn, Tseytlin:1999ii}.
    
    The emergence of an $\ads_3$ throat for a large number of $\text{D}1$-branes suggests a holographic interpretation of the results, opening new avenues in top-down holography with broken supersymmetry. Moreover, branes with the same charges always repel, corroborating the Weak Gravity Conjecture (WGC)~\cite{ArkaniHamed:2006dz} in a non-supersymmetric context. In the probe regime, this behavior stems from a novel renormalization of the effective charge-to-tension ratio due to the supersymmetry-breaking gravitational tadpoles~\cite{Antonelli:2019nar}, while in the world-volume gauge theory it stems from the absence of the fermionic contribution to the one-loop effective potential. More generally, the qualitative repulsive or attractive behavior is shared among different regimes whenever they apply. This non-trivial agreement points to a tantalizing deeper principle behind our findings.

\section{Branes and gravitational tadpoles}\label{sec:string_models}    

In this section we briefly review the three non-supersymmetric string models that we shall investigate on in this paper, focusing on their brane content. In particular, the $USp(32)$ model of~\cite{Sugimoto:1999tx} and the $U(32)$ type $0'$B model of~\cite{Sagnotti:1995ga, Sagnotti:1996qj} contain several charged and uncharged branes in their perturbative spectra~\cite{Dudas:2001wd}. In suitable probe regimes, their dynamics can be studied via string amplitudes or effective world-volume actions, while in the opposite regime their back-reaction can be described from the bulk gravitational action. Similarly, the heterotic NS5-branes in the $SO(16) \times SO(16)$ model of~\cite{AlvarezGaume:1986jb, Dixon:1986iz} can be studied at low energies, despite the absence of strong-weak dualities\footnote{See~\cite{Blum:1997cs, Blum:1997gw} for some efforts in this direction. A possible non-perturbative construction of the $SO(16) \times SO(16)$ heterotic model, along the lines of Ho\v{r}ava-Witten theory~\cite{Horava:1995qa}, has been proposed in~\cite{Faraggi:2007tj}. It would be interesting to describe NS5-branes in this setting, and I would like to thank A. Faraggi for pointing this out to me.}.

\subsection{Tachyon-free vacuum amplitudes}\label{sec:vacuum_amplitudes}

Let us begin constructing the relevant ten-dimensional string models starting from the one-loop vacuum amplitudes of their parent models. While the type $0'$B model arises as a non-tachyonic orientifold of the tachyonic type $0$B model, and is thus non-supersymmetric at the outset, the $USp(32)$ model and the $SO(16) \times SO(16)$ heterotic model arise as projections of the type IIB and the $E_8 \times E_8$ superstrings respectively, thereby featuring supersymmetry breaking at the string scale.

\subsubsection{The orientifold models}\label{sec:orientifolds}

Let us now introduce the relevant orientifold projections~\cite{Sagnotti:1987tw, Pradisi:1988xd, Horava:1989vt, Horava:1989ga, Bianchi:1990yu, Bianchi:1990tb, Bianchi:1991eu, Sagnotti:1992qw} via one-loop vacuum amplitudes\footnote{The following construction is based on the characters $\left( O_{2n} \, , V_{2n} \, , S_{2n} \, , C_{2n} \right)$ of the level-one affine $\mathfrak{so}(2n)$ algebra. See~\cite{Dudas:2000bn, Angelantonj:2002ct, Mourad:2017rrl} for reviews.}, starting from the more familiar case of the type I superstring. The torus amplitude
\begin{eqaed}\label{eq:torus_amp_i}
    \mathcal{T}_{\text{I}} = \frac{1}{2} \int_{\mathcal{F}} \frac{d^2\tau}{\tau_2^6} \, \frac{\left(V_8 - S_8 \right) \overline{\left(V_8 - S_8 \right)}}{|\eta(\tau)|^{16}}
\end{eqaed}
is halved with respect to its type IIB counterpart, and is to be added to the amplitudes pertaining to the open-string and unoriented sectors. These are associated to the Klein bottle, the annulus and the Möbius strip, and read
\begin{eqaed}\label{eq:klein_annulus_mobius_mod}
	\mathcal{K} & = \frac{1}{2} \int_0^\infty \frac{d\tau_2}{\tau_2^6} \, \frac{\left( V_8 - S_8 \right) \!\left(2i \tau_2\right)}{\eta^8\!\left(2i \tau_2 \right)} \, , \\
	\mathcal{A} & = \frac{N^2}{2} \int_0^\infty \frac{d\tau_2}{\tau_2^6} \, \frac{\left( V_8 - S_8 \right) \!\left(\frac{i\tau_2}{2}\right)}{\eta^8\!\left(\frac{i\tau_2}{2} \right)} \, , \\
	\mathcal{M} & = \frac{\varepsilon \, N}{2} \int_0^\infty \frac{d\tau_2}{\tau_2^6} \, \frac{\left( \widehat{V}_8 - \widehat{S}_8 \right) \!\left(\frac{i\tau_2}{2} + \frac{1}{2}\right)}{\widehat{\eta}^8\!\left(\frac{i\tau_2}{2} + \frac{1}{2} \right)} \, .
\end{eqaed}
The corresponding (loop-channel) UV divergences arise from tadpoles in the NS-NS and R-R sectors, whose cancellation requires
\begin{eqaed}\label{eq:tadpole_super_mod}
	N = 32 \, , \qquad \varepsilon = - 1 \, .
\end{eqaed}
Similarly, one can obtain the $USp(32)$ model of~\cite{Sugimoto:1999tx} introducing an $\text{O}9$-plane with positive tension and charge together with $\overline{\text{D}9}$-branes, yielding a vanishing R-R tadpole. However, the NS-NS tadpole is not canceled, and thus supersymmetry is broken at the string scale\footnote{More precisely, supersymmetry is preserved in the closed-string sector, but it is non-linearly realized in the open-string sector via ``brane supersymmetry breaking'' (BSB)~\cite{Antoniadis:1999xk, Angelantonj:1999jh, Aldazabal:1999jr, Angelantonj:1999ms}.}.

This amounts to changing a sign in the Möbius strip amplitude, which now takes the form
\begin{eqaed}\label{eq:mobius_bsb_mod}
	\mathcal{M}_{\text{BSB}} = \frac{\varepsilon \, N}{2} \int_0^\infty \frac{d\tau_2}{\tau_2^6} \, \frac{\left( \widehat{V}_8 + \widehat{S}_8 \right) \!\left(\frac{i\tau_2}{2} + \frac{1}{2}\right)}{\widehat{\eta}^8\!\left(\frac{i\tau_2}{2} + \frac{1}{2} \right)} \, .
\end{eqaed}
In order to cancel the R-R tadpole $\varepsilon = 1$ and $N = 32$, leading to a $USp(32)$ gauge group. Since the residual tension in the NS-NS tadpole does not cancel, the low-energy physics of this model includes the string-frame runaway exponential potential\footnote{See~\cite{Dudas:2000nv, Pradisi:2001yv} for more details on the low-energy couplings of BSB models.}
\begin{eqaed}\label{eq:runaway_potential_string_frame}
	T \int d^{10} x \, \sqrt{- g_{\rm s}} \, e^{- \phi} \, ,
\end{eqaed}
whose Einstein-frame counterpart is
\begin{eqaed}\label{eq:runaway_potential_einstein_frame}
	T \int d^{10} x \, \sqrt{- g} \, e^{\gamma \phi} \, , \qquad \gamma = \frac{3}{2} \, .
\end{eqaed}
Similar exponential potentials actually appear also in the other models at stake, and resonate with the considerations in~\cite{Buratti:2021yia}. We shall review some aspects of their dramatic back-reaction in the next section.

Starting from the type $0\text{B}$ model, which is described by the torus amplitude
\begin{eqaed}\label{eq:0_mod}
	\mathcal{T}_{\text{0B}} = \int_{\mathcal{F}} \frac{d^2\tau}{\tau_2^6} \, \frac{O_8 \, \overline{O_8} + V_8 \, \overline{V_8} + S_8 \, \overline{S_8} + C_8 \, \overline{C_8}}{|\eta(\tau)|^{16}} \, ,
\end{eqaed}
one can perform an orientifold projection to obtain the $U(32)$ type $0'$B model of~\cite{Sagnotti:1995ga, Sagnotti:1996qj}. This involves adding to (half of) eq.~\eqref{eq:0_mod} the open-sector amplitudes
\begin{eqaed}\label{eq:0'b_annulus_mobius_mod}
	\mathcal{K}_{0'\text{B}} & = \frac{1}{2} \int_0^\infty \frac{d\tau_2}{\tau_2^6} \left( - \, O_8 + V_8 + S_8 - C_8 \right) \, , \\
	\mathcal{A}_{0'\text{B}} & \int_0^\infty \frac{d\tau_2}{\tau_2^6} \, n \, \overline{n} \, V_8 - \, \frac{n^2 + \overline{n}^2}{2} \, C_8 \, , \\
	\mathcal{M}_{0'\text{B}} & = \int_0^\infty \frac{d\tau_2}{\tau_2^6} \, \frac{n + \overline{n}}{2} \, \widehat{C}_8 \, ,
\end{eqaed}
where tadpole cancellation fixes $n = \overline{n} = 32$ and a $U(32)$ gauge group. The corresponding O9-plane, which arises from a particular combination of elementary O9-planes~\cite{Angelantonj:2002ct}, has vanishing tension, and therefore the relevant exponential potential is halved with respect to eq.~\eqref{eq:runaway_potential_einstein_frame}.

\subsubsection{The heterotic model}\label{sec:heterotic}

As we have anticipated, the $E_8 \times E_8$ superstring admits a tachyon-free projection that breaks supersymmetry~\cite{AlvarezGaume:1986jb, Dixon:1986iz}. In particular, starting from the torus amplitude
\begin{eqaed}\label{eq:torus_amp_h}
    \mathcal{T}_{\text{HE}} = \int_{\mathcal{F}} \frac{d^2\tau}{\tau_2^6} \, \frac{\left(V_8 - S_8 \right) \overline{\left( O_{16} + S_{16} \right)}^2}{|\eta(\tau)|^{16}} \, ,
\end{eqaed}
one can project onto the states with even total fermion number\footnote{Let us recall that T-duality corresponds to a projection onto the states with even right-moving fermion number.}. Thus, one is to add to the projected torus amplitude its images under $S$ and $T$ modular transformations in order to restore modular invariance. The final amplitude reads
\begin{eqaed}\label{eq:h_soxso_mod}
	\mathcal{T}_{SO(16) \times SO(16)} = \int_{\mathcal{F}} \frac{d^2\tau}{\tau_2^6} \frac{1}{|\eta(\tau)|^{16}} \, & \bigg[  O_8 \, \overline{\left( V_{16} \, C_{16} + C_{16} \, V_{16} \right)} \\
	& + V_8 \, \overline{\left( O_{16} \, O_{16} + S_{16} \, S_{16} \right)} \\
	& - S_8 \, \overline{\left( O_{16} \, S_{16} + S_{16} \, O_{16} \right)} \\
	& - C_8 \, \overline{\left( V_{16} \, V_{16} + C_{16} \, C_{16} \right)} \bigg] \, .
\end{eqaed}
While tachyons are absent from the perturbative spectrum by virtue of level matching, the one-loop vacuum energy does not vanish\footnote{Some projections allow vanishing or suppressed leading contributions to the vacuum energy~\cite{Dienes:1990ij,Dienes:1990qh,Kachru:1998hd, Angelantonj:2004cm, Abel:2015oxa, Abel:2017rch}.}, and its value is of order one in string units. In the string-frame low-energy effective action it appears as a cosmological constant, and thus as a runaway exponential potential
\begin{eqaed}\label{eq:runaway_potential_het}
	T \int d^{10}x \, \sqrt{- g} \, e^{\gamma \phi} \, , \qquad \gamma = \frac{5}{2} 
\end{eqaed}
in the Einstein frame. All in all, the low-energy manifestation of gravitational tadpoles in both the orientifold models and in the $SO(16) \times SO(16)$ heterotic model can be encompassed by the same type of exponential potential for the dilaton.

\subsection{Charged and uncharged branes}\label{sec:brane_scan}

One-loop vacuum amplitudes can be also employed to investigate the D-brane content of the orientifold models, as described in detail in~\cite{Dudas:2001wd}. Indeed, the consistency of (transverse-channel) D$p$-D$p$ and D$p$-D9 annulus amplitudes and Möbius strip amplitudes in the R-R sectors allows one to recover the spectrum of charged ``BPS-like'' branes, while combining their contributions reconstructs the dimensions of the adjoint representations of the world-volume gauge group, namely $USp(2N)$ or $SO(2N)$. Finally, the coefficients in front of the NS-NS and R-R characters correspond to tension and charge respectively.

The analysis in~\cite{Dudas:2001wd} shows that, similarly to the type I superstring, the $USp(32)$ orientifold model contains charged D1-branes and D5-branes, whose world-volume gauge groups are symplectic and orthogonal respectively. Moreover, the remaining values of $p$ pertain to uncharged branes, whose stability can be addressed studying tachyonic excitations. In particular, a single D3-brane and a single D4-brane are free of tachyons in this model, while a single D0-brane, whose tachyons belong to the adjoint representation of $USp(2) \simeq SU(2)$, is unstable. Similarly, a single D2-brane, whose tachyons belong to the anti-symmetric (singlet) representation of $USp(2)$, is unstable. The other D$p$-branes are unstable on account of bi-fundamental tachyons, arising from the interaction with the background $\overline{\text{D}9}$-branes.

The spectrum of D-branes in the type $0'$B model was studied in~\cite{Dudas:2000sn, Dudas:2001wd}. The model contains charged D$p$-branes with $p$ odd, and while their world-volume excitations are devoid of tachyons, D$p$-D$q$ exchanges include tachyons for $\abs{p - q} < 4$, analogously to the type IIB setting. Furthermore, once again mirroring the type IIB setting, the world-volume gauge groups are unitary. However, the D9-D7 exchange spectrum contains a tachyon. Even values of $p$ pertain to uncharged branes.

At leading order the D$p$-D$p$ interaction between charged branes vanishes~\cite{Dudas:2001wd}, but one is to take into account the presence of the $\overline{\text{D}9}$-branes and O9-plane, which bring along non-trivial contributions. The main aim of this paper is to investigate this interaction in a number of complementary regimes, and the resulting dynamics appears to realize the WGC in a non-trivial fashion~\cite{Antonelli:2019nar} when supersymmetry is broken.

\subsection{Low-energy effective description}\label{sec:low-energy}

Let us now introduce the low-energy effective description pertaining to the models introduced in Section~\ref{sec:vacuum_amplitudes}. As we have anticipated, both the orientifold models of Section~\ref{sec:orientifolds} and the heterotic model of Section~\ref{sec:heterotic} can be described, at low energies, by an Einstein-frame action of the form\footnote{Throughout this paper we use the ``mostly plus'' metric signature.}
\begin{eqaed}\label{eq:action}
S = \frac{1}{2\kappa_D^2}\int d^D x \, \sqrt{-g} \, \left( R - \frac{4}{D-2} \left(\partial \phi \right)^2 - V(\phi) - \frac{f(\phi)}{2(p+2)!}\, H_{p+2}^2 \right) \, ,
\end{eqaed}
where the bosonic fields include a dilaton $\phi$ and a $(p+2)$-form field strength $H_{p+2} = dB_{p+1}$. In the relevant string models $D = 10$, while
\begin{eqaed}\label{eq:potential_form-coupling}
V(\phi) = T \, e^{\gamma \phi} \, , \qquad f(\phi) = e^{\alpha \phi} \, ,
\end{eqaed}
and their perturbative spectra also include Yang-Mills fields, whose contribution to the action takes the form
\begin{eqaed}\label{eq:gauge_action}
	S_{\text{gauge}} = - \, \frac{1}{2\kappa_D^2} \int d^D x \, \sqrt{-g} \left( \frac{w(\phi)}{4} \, \text{Tr} \, \mathcal{F}_{MN} \, \mathcal{F}^{MN} \right)
\end{eqaed}
with $w(\phi)$ an exponential. Although $\ads$ compactifications supported by gauge fields of this type were studied in~\cite{Mourad:2016xbk}, their perturbative corners seem not to exhibit any novel features, and thus we shall neglect this contribution to the EFT action in this paper.

The (bosonic) low-energy dynamics of the orientifold models that we have presented in Section~\ref{sec:orientifolds} is described by the Einstein-frame parameters
\begin{eqaed}\label{eq:bsb_electric_params}
D = 10 \, , \quad p = 1 \, , \quad \gamma = \frac{3}{2} \, , \quad \alpha = 1 \, ,
\end{eqaed}
while the residual tension
\begin{eqaed}\label{eq:T_bsb}
T = 2\kappa_{10}^2 \times 64 \, T_{\text{D}9} = \frac{16}{\pi^2 \, \alpha'}
\end{eqaed}
in the BSB model reflects the presence of $16$ $\overline{\text{D}9}$-branes and the O9-plane~\cite{Sugimoto:1999tx}. In the type $0'\text{B}$ model $T$ is half of this value, since the corresponding O9-plane is tensionless.

The heterotic model of Section~\ref{sec:heterotic} is described by the Einstein-frame parameters
\begin{eqaed}\label{eq:het_electric_params}
D = 10 \, , \quad p = 1 \, , \quad \gamma = \frac{5}{2} \, , \quad \alpha = -1 \, ,
\end{eqaed}
and the one-loop cosmological constant $T$, which was estimated in~\cite{AlvarezGaume:1986jb}, is of order one in string units. Dualizing the Kalb-Ramond form one can equivalently work with the parameters
\begin{eqaed}\label{eq:het_magnetic_params}
D = 10 \, , \quad p = 5 \, , \quad \gamma = \frac{5}{2} \, , \quad \alpha = 1 \, ,
\end{eqaed}
which highlight the electric coupling of NS5-branes to the dual $B_6$ potential.

Let us collect a few remarks on the reliability of the effective action of eq.~\eqref{eq:action}. The dilaton potential contains one less power of $\alpha'$ with respect to the other terms, and thus its runaway effects are to be balanced by an additional control parameter. The $\ads$ landscapes studied in~\cite{Mourad:2016xbk, Basile:2018irz, Antonelli:2019nar} achieve this by means of large fluxes, whereby curvature corrections and string loop corrections are expected to be under control. However, in the orientifold models these are R-R fluxes, and thus a world-sheet formulation appears subtle\footnote{It is worth noting that world-sheet CFTs on $\ads_3$ backgrounds have been related to $\alpha'$-exact WZW models~\cite{Maldacena:2000hw}, which in principle could be relevant in this case.}. In the heterotic model the fluxes are NS-NS, but the dilaton tadpole arises at one-loop level. On the other hand, the flux-less Dudas-Mourad solutions~\cite{Dudas:2000ff} contains regions where string-loop and curvature corrections are expected to be important.

The field equations stemming from the action in eq.~\eqref{eq:action} read
\begin{eqaed}\label{eq:eoms_eft}
	R_{MN} & = \widetilde{T}_{MN} \, , \\
	\Box \, \phi \, - V'(\phi) - \, \frac{f'(\phi)}{2(p+2)!} \, H_{p+2}^2 & = 0 \, , \\
	d \star (f(\phi) \, H_{p+2}) & = 0 \, ,
\end{eqaed}
where the (trace-reversed) stress-energy tensor
\begin{eqaed}\label{eq:trace-reversed_stress}
	\widetilde{T}_{MN} \equiv T_{MN} - \frac{1}{D-2} \, {T^A}_A \, g_{MN} \, ,
\end{eqaed}
and we use conventions such that
\begin{eqaed}\label{eq:stress_definition}
	T_{MN} \equiv - \, \frac{\delta S_{\text{matter}}}{\delta g^{MN}} \, .
\end{eqaed} 
For the action in eq.~\eqref{eq:action}, one arrives at
\begin{eqaed}\label{eq:stress_tensor}
	\widetilde{T}_{MN} = \; & \frac{4}{D-2} \, \partial_M \phi \, \partial_N \phi + \frac{f(\phi)}{2(p+1)!} \left(H_{p+2}^2\right)_{MN} \\
	& + \frac{g_{MN}}{D-2} \left( V - \frac{p+1}{2(p+2)!} \, f(\phi) \, H_{p+2}^2 \right) \, ,
\end{eqaed}
where $\left(H_{p+2}^2\right)_{MN} \equiv H_{M A_1 \dots A_{p+1}} \, {H_N}^{A_1 \dots A_{p+1}}$.

\section{Back-reaction of non-supersymmetric branes}\label{sec:geometries}

Let us now turn to the gravitational back-reaction of the branes described in Section~\ref{sec:brane_scan}. The motivation to investigate it is two-fold: on one hand, we would like to understand interactions between two stacks of branes beyond the regime in which both stacks are light. In particular, whenever one of the stacks is parametrically heavier one can expect to describe the interaction replacing the heavy stack with its gravitational back-reaction. On the other hand, large fluxes appear to constitute the most reliable tool to retain control of the low-energy EFT in the presence of gravitational tadpoles~\cite{Mourad:2016xbk}, and the (near-horizon limit of the) geometries generated by charged branes can realize settings of this type in a natural fashion~\cite{Antonelli:2019nar}.

\subsection{Static Dudas-Mourad solutions}\label{sec:static_solutions}

The Dudas-Mourad solutions of~\cite{Dudas:2000ff} comprise static solutions with nine-dimensional Poincar\'e symmetry\footnote{A T-dual version of this configuration in the $USp(32)$ model was investigated in~\cite{Blumenhagen:2000dc}.}, where one dimension is compactified on an interval, and ten-dimensional cosmological solutions. In the ensuing discussion we shall focus on the former, since, at least in the orientifold models, they appear to afford an interpretation in terms of D8-branes~\cite{Antonelli:2019nar}.

The gravitational tadpole signals the absence of a ten-dimensional Minkowski solution, and therefore the maximal possible symmetry available to static solutions is nine-dimensional Poincar\'e symmetry. Correspondingly, the most general solution of this type is a warped product of nine-dimensional Minkowski space-time, parametrized by coordinates $x^\mu$, and a one-dimensional internal space, parametrized by a coordinate $y$. In the orientifold models of Section~\ref{sec:orientifolds}, the Einstein-frame solution is given by
\begin{eqaed}\label{eq:dm_orientifold_einstein_mod}
	ds_\text{orientifold}^2 & = \abs{\alpha_\text{O} \, y^2}^{\frac{1}{18}} \, e^{- \frac{\alpha_\text{O} y^2}{8}} \, dx_{1,8}^2 + e^{- \frac{3}{2} \Phi_0} \, \abs{\alpha_\text{O} \, y^2}^{- \frac{1}{2}} \, e^{- \frac{9 \alpha_\text{O} y^2}{8}} dy^2 \, , \\
	\phi & = \frac{3}{4} \, \alpha_\text{O} \, y^2 + \frac{1}{3} \, \log \abs{\alpha_\text{O} \, y^2} + \Phi_0 \, ,
\end{eqaed}
where, here and in the following,
\begin{eqaed}\label{eq:minkowski_p_mod}
	dx_{1,p}^2 \equiv \eta_{\mu \nu} \, dx^\mu \, dx^\nu
\end{eqaed}
denotes the $(p+1)$-dimensional Minkowski metric. The absolute values in eq.~\eqref{eq:dm_orientifold_einstein_mod}, as well as the singularities in the dilaton profile, highlight that the physical range of the internal coordinate is $y \in (0,\infty)$. In the heterotic model of Section~\ref{sec:heterotic} the Einstein-frame solution is given by
\begin{eqaed}\label{eq:dm_heterotic_einstein_mod}
	ds_\text{heterotic}^2 = \, & \left(\sinh \abs{\sqrt{\alpha_\text{H}} \, y}\right)^{\frac{1}{12}} \left(\cosh \abs{\sqrt{\alpha_\text{H}} \, y}\right)^{- \frac{1}{3}} dx_{1,8}^2 \\
	& + e^{- \frac{5}{2} \Phi_0} \left(\sinh \abs{\sqrt{\alpha_\text{H}} \, y}\right)^{- \frac{5}{4}} \left(\cosh \abs{\sqrt{\alpha_\text{H}} \, y}\right)^{-5} dy^2 \, , \\
	\phi = & \, \frac{1}{2} \, \log \, \sinh \abs{\sqrt{\alpha_{\text{H}}} \, y} + 2 \, \log \, \cosh \abs{\sqrt{\alpha_{\text{H}}} \, y} + \Phi_0 \, ,        
\end{eqaed}
and similar considerations on the range of $y$ apply.

In accord with the conventions of~\cite{Dudas:2000ff}, in eqs.~\eqref{eq:dm_orientifold_einstein_mod} and~\eqref{eq:dm_heterotic_einstein_mod} the scales $\alpha_\text{O,H} \equiv \frac{T}{2}$, while $\Phi_0$ is a free parameter which specifies the value of the local string coupling at some $y = y_0$. In both solutions, the internal spaces parametrized by $y$ are actually intervals of finite length
\begin{eqaed}\label{eq:dm_length}
    R_c \equiv \int_0^\infty \sqrt{g_{yy}} \, dy < \infty \, ,
\end{eqaed}
and, for $g_s \equiv e^{\Phi_0} \ll 1$, the interior of the parametrically wide interval is weakly coupled. Moreover, since $\alpha_\text{O,H} \propto T$, as one approaches the supersymmetric case the internal length diverges\footnote{Let us remark that, strictly speaking, $T$ cannot be sent to zero, but it is still instructive to consider the formal limit $T \to 0$.}.

As described in~\cite{Antonelli:2019nar}, at least in the orientifold models the Dudas-Mourad solution resonates with the back-reaction of D8-branes. Indeed, the back-reaction of a stack of parallel branes in the EFT described by eq.~\eqref{eq:action} can be described via a Toda-like reduced dynamical system, and the $p=8$ case yields an integrable system.

\subsection{Back-reaction of extremal branes}\label{sec:extremal_reduced_dynamical_system}

In this section we review the construction of the Toda-like reduced dynamical system for extremal branes, outlining the resulting gravitational back-reaction~\cite{Antonelli:2019nar}. Extremal $p$-branes entail a residual $SO(1,p) \times SO(q)$ space-time symmetry, so that in a suitable gauge the most general solution to eq.~\eqref{eq:eoms_eft} takes the form
\begin{eqaed}\label{eq:brane_full_ansatz}
    ds^2 & = e^{\frac{2}{p+1}v - \frac{2q}{p}b} \, dx^2_{1,p} + e^{2v-\frac{2q}{p}b} \, dr^2 +e^{2b} \, R^2_0 \, d\Omega_q^2 \, , \\
    \phi & = \phi(r) \, , \\
    H_{p+2} & = \frac{n}{f(\phi)(R_0 \, e^b)^q} \, \Vol_{p+2} \, , \qquad \Vol_{p+2} = e^{2v - \frac{q}{p}(p+2)b} \, d^{p+1} x\,\wedge\, dr \, ,
\end{eqaed}
where $r$ is the transverse radial coordinate, $R_0$ is an arbitrary reference radius. The resulting equations stem from a constrained Toda-like system~\cite{Klebanov:1998yya,Dudas:2000sn}, described by the action
\begin{eqaed}\label{eq:toda_action}
    S_{\text{red}} = \int dr \left[ \frac{4}{D-2} \left(\phi'\right)^2 - \frac{p}{p+1} \left(v'\right)^2 + \frac{q(D-2)}{p} \left(b'\right)^2 - \, U \right] \, ,
\end{eqaed}
with
\begin{eqaed}\label{eq:toda_potential}
    U = - \,  T \, e^{\gamma \phi + 2v - \frac{2q}{p}b} - \frac{n^2}{2R_0^{2q}} \, e^{-\alpha\phi + 2v-\frac{2q(p+1)}{p}b} + \frac{q(q-1)}{R_0^2} \, e^{2v-\frac{2(D-2)}{p}b} \, ,
\end{eqaed}
and the Hamiltonian constraint reads
\begin{eqaed}\label{eq:toda_constraint}
    \frac{4}{D-2} \left(\phi'\right)^2 - \, \frac{p}{p+1} \left(v'\right)^2 + \frac{q(D-2)}{p} \left(b' \right)^2 + U = 0 \, .
\end{eqaed}
Here, the (electric) flux $n$ ought to be proportional to the number $N$ of branes, and is defined by
\begin{eqaed}\label{eq:electric_flux}
n = \frac{1}{\Omega_q} \int_{\ess^q} f \star H_{p+2} = c \, e^{\alpha \phi} \, R^q
\end{eqaed}
with $\Omega_q$ the volume of the unit $q$-sphere.

\begin{figure}
    \centering
    \scalebox{0.9}{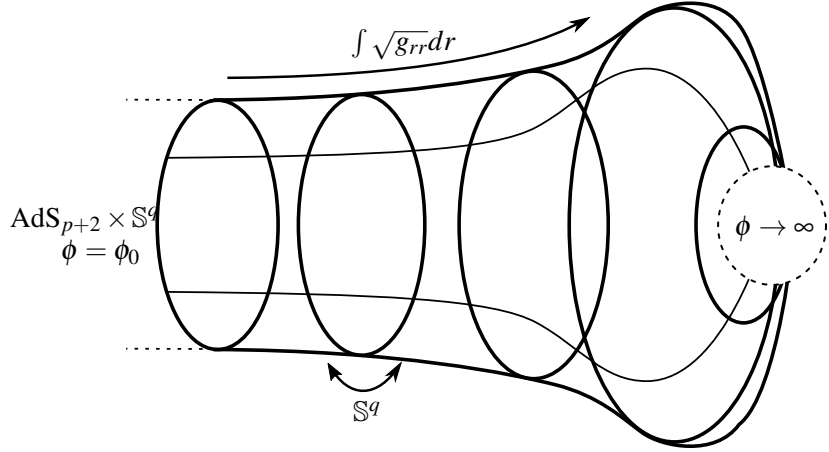}
    \caption{a schematic depiction of the back-reacted geometry sourced by the branes, where the world-volume directions are excluded. The geometry interpolates between the $\adsts$ throat and the pinch-off singularity.}
    \label{fig:geometry}
\end{figure}

In~\cite{Antonelli:2019nar} the $\ads \times \ess$ solutions of~\cite{Mourad:2016xbk} were recovered according to
\begin{eqaed}\label{eq:ads_s_toda}
    \phi & = \phi_0 \, , \\
    e^v & = \frac{L}{p+1} \, \left(\frac{R}{R_0}\right)^{\frac{q}{p}} \, \frac{1}{-r} \, , \\
    e^b & = \frac{R}{R_0} \, ,
\end{eqaed}
where $r < 0$. Anticipating that this solution arises as a near-horizon limit, this choice places the core at $r \; \to \; -\infty$. Rescaling $x$ by a constant, and substituting
\begin{eqaed}\label{eq:toda_ads_s_diffeo}
    r \; \mapsto \; - \, \frac{z^{p+1}}{p+1}
\end{eqaed}
in the metric of eq.~\eqref{eq:brane_full_ansatz} reveals that the solution in eq.~\eqref{eq:ads_s_toda} is indeed $\adsts$ in a Poincar\'e patch. Following the analogy with supersymmetric cases, where infinite $\ads$ throats behave as attractors, in~\cite{Antonelli:2019nar} it was shown that radial perturbations always contain modes that decay going toward the horizon, while the blow-up modes are expected to be associated to extremality breaking. While control of the full solution appears necessary to connect the asymptotic parameters to these modes, one can verify that they reassuringly match in number studying the sub-leading behavior in the far-away region. Indeed, while we shall not need this result in the following, it turns out that the back-reacted geometry ends, \emph{at a finite radial geodesic distance}, in a pinch-off singularity where strong-coupling effects are expected to play a crucial rôle. The resulting suggestive, albeit incomplete, picture is presented in fig.~\ref{fig:geometry}, and highlights the obstructions in defining tension and flux as asymptotic quantities.

\subsubsection{Non-perturbative $\ads$ instabilities}\label{sec:bubbles}

The $\ads \times \ess$ near-horizon solution of eq.~\eqref{eq:ads_s_toda} can be recast in the coordinate-free fashion
\begin{eqaed}\label{eq:adsxs_ansatz}
	ds^2 & = L^2 \, ds_{\ads_{p+2}}^2 + R^2 \, d\Omega_q^2 \, , \\
	H_{p+2} & = c \, \Vol_{\ads_{p+2}} \, , \\
	\phi & = \phi_0 \, ,
\end{eqaed}
where $ds_{\ads_{p+2}}^2$ is the unit-radius space-time metric and $\Vol_{\ads_{p+2}}$ denotes the canonical volume form on $\ads_{p+2}$ with radius $L$. The geometry exists if and only if the parameters $\gamma \, , \alpha$ in eq.~\eqref{eq:potential_form-coupling} satisfy
\begin{eqaed}\label{eq:adsxs_existence_conditions}
\alpha > 0 \, , \qquad q > 1 \, , \qquad \left(q-1\right) \gamma > \alpha \, ,
\end{eqaed}
and using eq.~\eqref{eq:potential_form-coupling} the string coupling $g_s = e^{\phi_0}$ and the curvature radii $L \, , \, R$ are given by
\begin{eqaed}\label{eq:ads_s_solution}
	c & = \frac{n}{g_s^\alpha R^q} \, , \\
	g_s^{(q-1)\gamma-\alpha} & = \left(\frac{(q-1)(D-2)}{\left(1+\frac{\gamma}{\alpha} \left(p+1 \right) \right)T}\right)^q \,\frac{2\gamma T}{\alpha n^2} \, , \\
	R^{2\frac{\left(q-1\right)\gamma-\alpha}{\gamma}} & = \left( \frac{\alpha + \left(p+1\right) \gamma}{(q-1)(D-2)}\right)^{\frac{\alpha+\gamma}{\gamma}} \left(\frac{T}{\alpha}\right)^{\frac{\alpha}{\gamma}}\frac{n^2}{2\gamma} \, , \\
	L^2 & = R^2 \left(\frac{p+1}{q-1} \cdot \frac{\left(p+1\right)\gamma+ \alpha}{\left(q-1\right)\gamma- \alpha}\right) \, .
\end{eqaed}
The solution of eq.~\eqref{eq:ads_s_solution}, originally found in~\cite{Mourad:2016xbk} for the string models presented in Section~\ref{sec:string_models}, was generalized and studied in detail in~\cite{Basile:2018irz, Antonelli:2019nar}. Among its intriguing features, it appears complementary to the supersymmetric $\ads_5 \times \ess^5$ solution of type IIB supergravity, but it has no moduli. Moreover, the large-$n$ limit corresponds both to small string couplings and small curvatures, and thus one can expect that the EFT description encoded in eqs.~\eqref{eq:bsb_electric_params} and~\eqref{eq:het_magnetic_params} be reliable in this regime. However, let us emphasize that a dimensionally reduced EFT description would not be consistent, since there is no scale separation. This is consistent with general considerations on scale separation~\cite{Lust:2019zwm, Lust:2020npd}\footnote{See also~\cite{DeLuca:2021mcj} for a detailed study of bounds on scale separation in flux compactifications.}, while the behavior of Kaluza-Klein excitations corroborates the Swampland distance conjecture (SDC)~\cite{Ooguri:2016pdq, Ooguri:2018wrx} and $\ads$ distance conjecture (ADC)~\cite{Lust:2019zwm}. The connection between the non-supersymmetric backgrounds that we have discussed and Swampland conjectures, in particular those regarding $\ds$ vacua~\cite{Obied:2018sgi, Ooguri:2018wrx, Garg:2018reu}, has been articulated in detail in~\cite{Basile:2020mpt}.

In light of the preceding discussion, the $\ads \times \ess$ solutions of eq.~\eqref{eq:ads_s_solution} appear to capture the near-horizon back-reaction of extremal branes in the presence of gravitational tadpoles. Specifically, they pertain to $\text{D}1$-branes in the orientifold models and $\text{NS}5$-branes in the heterotic model, while an analogous solution describing extremal $\text{D}5$-branes is still lacking and presents some subtleties\footnote{This state of affairs mirrors, to a certain extent, the peculiar near-horizon geometry sourced by BPS $\text{D}5$-branes, since it breaks the pattern of conformal $\ads_{p+2} \times \ess^{8-p}$ throats~\cite{Horowitz:1991cd, Stelle:1998xg, Boonstra:1998mp, Aharony:1999ti}.}. Although the internal sphere ought to reflect the simplest setting of parallel branes in the vacuum, it brings along perturbative instabilities~\cite{Basile:2018irz}. Since eq.~\eqref{eq:ads_s_solution} readily generalizes to any compact Einstein manifold, in principle it could be possible to eliminate the unstable modes choosing a different internal manifold, possibly pertaining to branes on conical singularities, and for the heterotic model an antipodal $\mathbb{Z}_2$ orbifold of the internal $\ess^3$ indeed achieves this. In this paper we shall thus neglect these instabilities.

However, these solutions are also fraught with non-perturbative instabilities, and undergo flux tunneling~\cite{Antonelli:2019nar}\footnote{For more details, see~\cite{Brown:1987dd,Brown:1988kg,Maldacena:1998uz,Seiberg:1999xz, BlancoPillado:2009di, Brown:2010bc, Brown:2010mg}.} in accord with the non-supersymmetric $\ads$ Swampland conjecture~\cite{Ooguri:2016pdq}. In particular, in~\cite{Antonelli:2019nar} it was shown that the semi-classical decay rate per unit volume of the $\ads$ solutions is given by $e^{-S_p^E}$, where the (extremized) instanton action of a brane with (bare) tension $T_p$ and charge $\mu_p$ is
\begin{eqaed}\label{eq:decay_rate_p}
    S_p^E & = T_p \, L^{p+1} \, g_s^{- \frac{\alpha}{2}} \, \Omega_{p+1} \, \mathcal{B}_p\left(v_0 \, \frac{\mu_p}{T_p}\right) \\
    & \propto n^{\frac{(p+1)\gamma+\alpha}{(q-1)\gamma-\alpha}} \, ,
\end{eqaed}
where
\begin{eqaed}\label{eq:Bp_function}
    \mathcal{B}_p(\beta) \equiv \frac{1}{(\beta^2 - 1)^{\frac{p+1}{2}}} - \frac{p+1}{2} \, \beta \, \int_0^{ \frac{1}{\beta^2 - 1}} \frac{u^{\frac{p}{2}}}{\sqrt{1+u}} \, du
\end{eqaed}
and
\begin{eqaed}\label{eq:v0_parameter}
    v_0 \equiv \sqrt{\frac{2(D-2)\gamma}{(p+1) ((q-1)\gamma-\alpha)}} \, .
\end{eqaed}
Remarkably, the tunneling process is allowed also in the extremal case $\mu_p = T_p$, since the charge-to-tension renormalization $v_0 > 1$ in the string models at stake. Indeed, from eq.~\eqref{eq:v0_parameter} one finds
\begin{eqaed}\label{eq:v0_orientifold}
    (v_0)_{\text{orientifold}} = \sqrt{\frac{3}{2}}
\end{eqaed}
for the orientifold models, and
\begin{eqaed}\label{eq:v0_heterotic}
    (v_0)_{\text{heterotic}} = \sqrt{\frac{5}{3}}
\end{eqaed}
for the heterotic model. Furthermore, consistency with the semi-classical limit requires that the string-frame tension $\tau_p = T_p \, e^{- \sigma \phi}$ of the brane scale with the dilaton according to 
\begin{eqaed}\label{eq:sigma_value}
    \sigma = \frac{2(p+1)}{D-2} + \frac{\alpha}{2} = 1 + \frac{\alpha_S}{2} \, ,
\end{eqaed}
where $\alpha_S$ is the string-frame counterpart of the parameter $\alpha$ in eq.~\eqref{eq:potential_form-coupling}. This result remarkably reproduces the correct couplings of fundamental branes, namely D-branes in the orientifold models, NS5-branes in the heterotic model and the ``exotic'' branes of~\cite{Bergshoeff:2005ac,Bergshoeff:2006gs,Bergshoeff:2011zk,Bergshoeff:2012jb,Bergshoeff:2015cba}, thereby suggesting that the branes sourcing the $\ads$ throat and nucleating in it are indeed D1-branes in the orientifold models and NS5-branes in the heterotic model. As we shall see shortly, the fact that $v_0 > 1$ also entails a novel realization of the WGC in the presence of gravitational tadpoles.

\subsection{D3-branes in the type \texorpdfstring{$0'$}{0'}B model}\label{sec:d3-branes}

As we have discussed, the $\ads \times \ess$ solutions that we have reviewed in the preceding section appear in some sense complementary to the supersymmetric type IIB setting. Indeed, for $p \neq 3 \, , \, 5$ the near-horizon geometry of BPS black branes in ten dimensions is conformal to $\ads_{p+2}$ with a singular warp factor~\cite{Horowitz:1991cd, Stelle:1998xg, Boonstra:1998mp, Aharony:1999ti}, while for $p = 3$ (and for M-branes in eleven-dimensional supergravity) it reproduces the familiar $\ads_5$. On the other hand, in the models that we have presented in Section~\ref{sec:string_models} there are no such solutions, since $\alpha = 0$ for D3-branes in the type $0'\text{B}$ model. The relevant near-horizon geometry that was studied in~\cite{Angelantonj:1999qg, Angelantonj:2000kh,Dudas:2000sn} actually involves an O3-plane, but its contribution is sub-leading for large fluxes. In~\cite{Angelantonj:2000kh} the authors found non-homogeneous deviations from $\ads_5 \times \mathbb{RP}^5$ which are suppressed, but not uniformly so, in the large-flux limit\footnote{Similar results in tachyonic type $0$ strings were obtained in~\cite{Klebanov:1998yya}.}. In detail, in coordinates in which the (string-frame) metric takes the form\footnote{The local expression in eq.~\eqref{eq:d3-branes_metric} does not account for the global distinction between $\ess^5$ and $\mathbb{RP}^5$.}~\cite{Angelantonj:2000kh}
\begin{eqaed}\label{eq:d3-branes_metric}
    ds^2 = R^2(u) \, \frac{du^2}{u^2} + \frac{\alpha'^2 \, u^2}{R^2(u)} \, dx^2_{1,3} + \widetilde{R}^2(u) \, d\Omega_5^2 \, ,
\end{eqaed}
the would-be $\ads_5$ and $\mathbb{RP}^5$ curvature radii $R(u) \, , \widetilde{R}(u)$ and the dilaton $\phi(u)$ acquire a dependence on the energy scale $u$ that, in the large-flux limit, behaves as
\begin{eqaed}\label{eq:d3-branes_correction}
    \frac{R^2(u)}{R^2_{\infty}} & \sim 1 - \frac{3}{16} \, g_s \, \alpha' T \, \log \left(\frac{u}{u_0}\right) \, , \\
    \frac{\widetilde{R}^2(u)}{R^2_{\infty}} & \sim 1 - \frac{3}{16 \sqrt[4]{8}} \, g_s^2 \, N \, \alpha' T \, \log \left(\frac{u}{u_0}\right) \, , \\
    \frac{1}{N} \, e^{-\phi} & \sim \frac{1}{g_s \, N} + \frac{3}{8 \sqrt[4]{8}} \, g_s \, \alpha' T \, \log \left(\frac{u}{u_0}\right) \, ,
\end{eqaed}
where $u_0$ is a reference scale, $R^2_\infty = \sqrt{4\pi \, g_s \, N}$ is the supersymmetric value of the radii and $N \gg 1$ ought to be interpreted as the number of D3-branes sourcing the geometry. In the large-$N$ limit the 't Hooft coupling $\lambda = 4\pi \, g_s \, N$ in the absence of the tadpole $T$ ought to be fixed, but the validity of the EFT description also requires $\lambda \gg 1$~\cite{Maldacena:1997re}, while on account of the second of eq.~\eqref{eq:d3-branes_correction} $g_s^2 \, N \ll 1$.

Although the resulting geometry is not completely under control because of the non-uniform character of these corrections, it is in principle amenable to numerical investigation via the Toda-like formalism that we have developed in Section~\ref{sec:extremal_reduced_dynamical_system}~\cite{Dudas:2000sn}. In the following section we shall discuss the interactions of D3-branes in the type $0'$B model in the regime in which a heavy stack generates the solution of eq.~\eqref{eq:d3-branes_correction}, and probe light stacks are subjects to a potential encoding the interaction.

\section{Brane interactions and the WGC}\label{sec:probes}

\begin{figure}[ht]
\centering
\includegraphics[width=.7\textwidth]{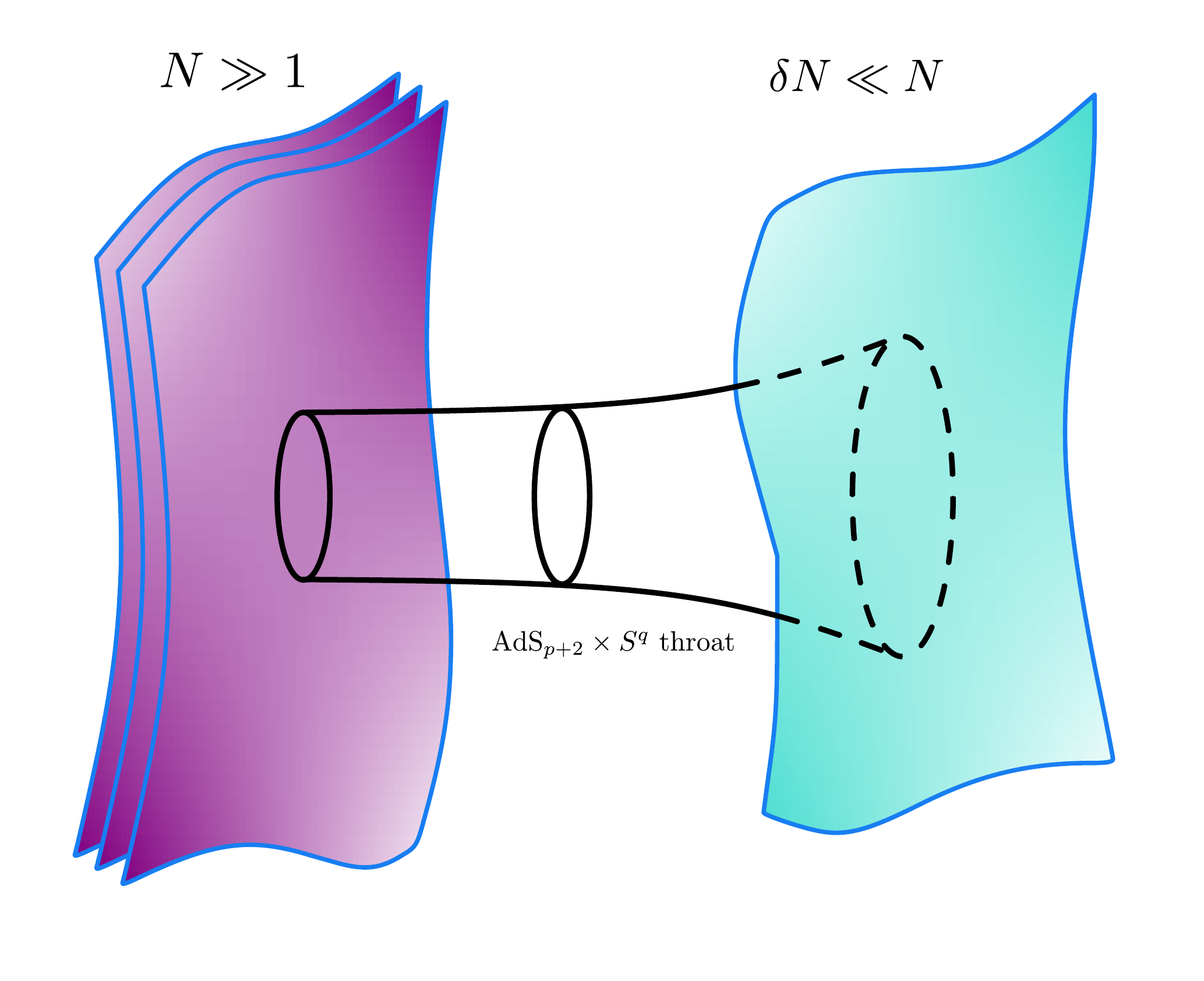}
\caption{a depiction of the interaction between a heavy stack of $N \gg 1$ branes and $\delta N<< N$ probe branes. The heavy stack sources the $\ads \times \ess$ throat probed by the light stack.}
\label{fig:throat_branes}
\end{figure}

In this section we study in detail the interactions between the branes that we have introduced in Section~\ref{sec:brane_scan}. To this end, we shall discuss a number of complementary regimes in which computations are expected to be under control. Namely, considering two parallel stacks of $N_p$ D$p$-branes and $N_q$ D$q$-branes, the cases that we shall address are the following:

\begin{itemize}
    \item The \emph{probe regime} $N_p \gg N_q$, in which one can replace the heavy stack of $N_p$ D$p$-branes with the corresponding back-reacted geometry probed by the D$q$-branes, as shown in fig.~\ref{fig:throat_branes}. We shall distinguish three cases: extremal branes probing $\ads$ throats, uncharged D$8$-branes probing $\ads$ throats, D$p$-branes probing the Dudas-Mourad geometry sourced by D$8$-branes. We shall also address NS5-branes probing the heterotic Dudas-Mourad geometry for completeness, although in that case an interpretation in terms of $8$-branes appears obscure in the absence of established dualities.
    
    \item The \emph{string-amplitude regime} $N_p \, , N_q = \mathcal{O}(1)$, which ought to be described by perturbative string amplitudes. In particular, since interactions between extremal branes yield vanishing annulus amplitudes, we shall focus on the case in which at least one stack is uncharged\footnote{The first non-trivial contribution to the potential between two extremal stacks would involve three-legged pants diagrams. In the bosonic case, the systematics of such computations were developed in~\cite{Bianchi:1988fr}.}.
    
    \item The \emph{holographic regime}, in which we shall consider the asymptotically free world-volume gauge theory that describes D1-branes. In this paper we shall focus on the weakly coupled UV regime, although the strongly coupled IR regime ought to be directly related to near-horizon $\ads$ throats. We shall discuss this interesting case in future work.
\end{itemize}

Despite the absence of supersymmetry, whenever any two regimes overlap we shall find qualitative agreement. In particular, extremal branes of equal dimension strictly repel, realizing the WGC in the absence of supersymmetry, while the NS-NS interactions in the presence of at least one uncharged stack are repulsive or attractive depending on the values of $p$ and $q$.

\subsection{Probe potentials and Weak Gravity}\label{sec:probe_potentials}

Let us begin analyzing the probe regime, replacing the heavy brane stack with its back-reacted geometry. In particular we shall focus on the near-horizon $\ads \times \ess$ throats that we have described in Section~\ref{sec:bubbles}, which ought to pertain to D1-branes and NS5-branes, and on the Dudas-Mourad geometry, which appears to arise from 8-branes. In order to encompass all the relevant cases, we shall consider a string-frame world-volume action of the form
\begin{eqaed}\label{eq:world-volume_action}
	S_p = - \, T_p \int d^{p+1} \zeta \, \sqrt{-j^* g_S} \, e^{- \sigma \phi} + \mu_p \int B_{p+1} \, ,
\end{eqaed}
where $j$ is the embedding of the world-volume coordinates $\zeta$ in space-time. Its Einstein-frame expression reads
\begin{eqaed}\label{eq:einstein-frame_actions_brane}
	S_p = - \, T_p \int d^{p+1} \zeta \, \sqrt{-j^* g} \, e^{\left(\frac{2(p+1)}{D-2} - \sigma \right) \phi} + \mu_p \int B_{p+1} \, ,
\end{eqaed}
and $\sigma = 1 \, , 2$ for D-branes and NS5-branes respectively. For the sake of generality we shall not assume that $T_p = \mu_p$ in the non-supersymmetric models at stake, although the results of~\cite{Dudas:2001wd} point toward extremality. However, as we shall now see in detail, in the string models that we have presented in Section~\ref{sec:string_models} even in this case charged brane stacks repel, due to a non-trivial renormalization of the charge-to-tension ratio mediated by supersymmetry breaking.

\subsubsection{Repulsive forces between extremal branes}\label{sec:probe_extremal}

Let us begin our analysis of probe-brane interactions considering the dynamics of an extremal $p$-brane moving in the $\ads_{p+2} \times \ess^q$ geometry of eq.~\eqref{eq:adsxs_ansatz}. The action in eq.~\eqref{eq:einstein-frame_actions_brane} encompasses both $\text{D}1$-branes in the orientifold models, where $B_2$ is a R-R potential, and $\text{NS}5$-branes in the heterotic model, where $B_6$ is the (magnetic dual of the) Kalb-Ramond potential. As we have stressed in the preceding section, the putative electromagnetic dual of these background would describe $\text{D}5$-branes and $\text{F}1$-strings, but their construction has proven elusive so far.

Let us remark that the dynamics at stake emerge spontaneously on account of the considerations in Section~\ref{sec:bubbles}, since bubble nucleation entails separation of pairs of branes and anti-branes. The results in~\cite{Antonelli:2019nar} show that like-charge branes are repelled while anti-branes are attracted, leading to brane-flux annihilation. In order to appreciate this, it is convenient to work in Poincar\'e coordinates, where the Einstein-frame metric of the $\ads \times \ess$ throat reads
\begin{eqaed}\label{eq:poincare_metric_adsxs}
    ds^2 = \frac{L^2}{z^2} \left(dz^2 + dx^2_{1,p} \right) + R^2 \, d\Omega^2_q \, ,
\end{eqaed}
choosing the world-volume embedding
\begin{eqaed}\label{eq:brane_embedding}
    j \, : \quad x^\mu = \zeta^\mu \, , \qquad z = Z(\zeta) \, , \qquad \theta^i = \theta^i_0
\end{eqaed}
with $\theta_0^i$ fixed coordinates on $\ess^q$. The action of eq.~\eqref{eq:einstein-frame_actions_brane} then evaluates to
\begin{eqaed}\label{eq:brane_action_adsxs}
    S_p & = - \tau_p \int d^{p+1}\! \zeta \, \left(\frac{L}{Z}\right)^{p+1} \left[ \sqrt{ 1 + \eta^{\mu\nu} \, \partial_\mu Z \, \partial_\nu Z } - \frac{c \, L}{p+1} \, \frac{\mu_p}{\tau_p}\right] \, ,
\end{eqaed}
where the dressed tension
\begin{eqaed}\label{eq:probe_dressed_tension}
    \tau_p \equiv T_p \, g_s^{- \frac{\alpha}{2}}
\end{eqaed}
on account of the considerations of Section~\ref{sec:bubbles}. Therefore, rigid branes are subject to the potential

\begin{eqaed}\label{eq:brane_potential}
    V_{\text{probe}}(Z) & = \tau_p \, \left(\frac{L}{Z}\right)^{p+1} \left[1 - \frac{c \, L \, g_s^{\frac{\alpha}{2}}}{p+1} \, \frac{\mu_p}{T_p} \right] \\
    & = \tau_p \, \left(\frac{L}{Z}\right)^{p+1} \left[1 - v_0 \, \frac{\mu_p}{T_p} \right] \, .
\end{eqaed}

\begin{figure}[ht]
	\centering
	\includegraphics[width=.7\textwidth]{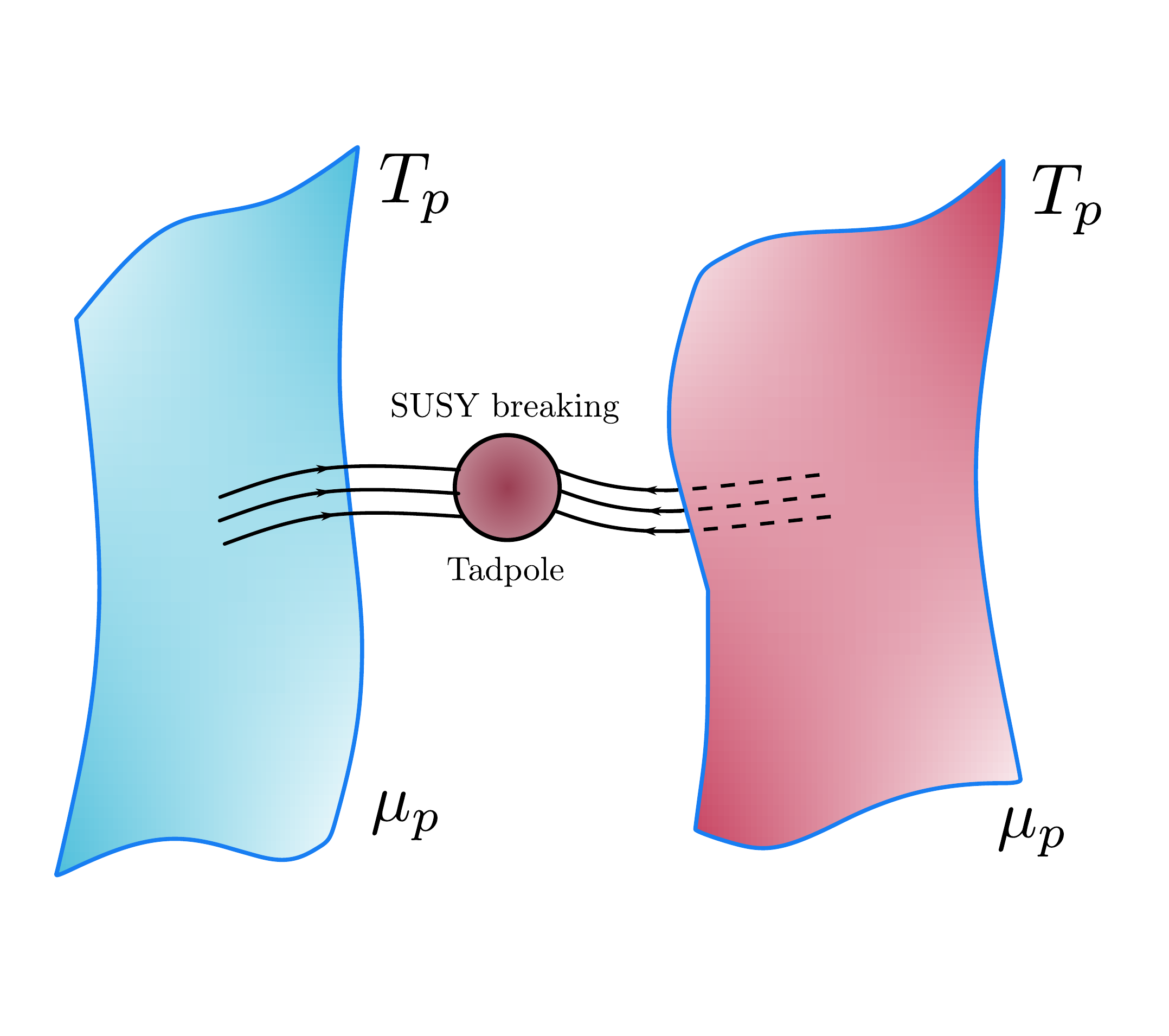}
	\caption{a depiction of the interaction between extremal branes mediated by supersymmetry breaking, reflecting the renormalization of the effective charge-to-tension ratio of eqs.~\eqref{eq:v0_parameter},~\eqref{eq:v0_orientifold} and~\eqref{eq:v0_heterotic}.}
	\label{fig:parallel_branes}
\end{figure}

While non-rigid branes exhibit richer dynamics~\cite{Antonelli:2019nar}, in the present setting we would like to emphasize that, as one can observe from eqs.~\eqref{eq:v0_orientifold} and~\eqref{eq:v0_heterotic}, extremal probes $\mu_p = T_p$ are \emph{indeed repelled by the stack}, being driven toward $Z \to 0$. In the orientifold models the picture is intuitive: $\text{D}1$-branes are mutually BPS\footnote{More precisely, despite the absence of supersymmetry, in the type $0'$B model charged branes exhibit a ``BPS-like'' no-force behavior at tree-level~\cite{Dudas:2001wd}.}, but their interaction with the space-time-filling $\overline{\text{D}9}$-branes and O9-plane renormalizes the charge-to-tension ratio, as depicted in fig.~\ref{fig:parallel_branes}. Similarly, in the heterotic model the corresponding interactions are mediated by the quantum-corrected vacuum energy. This peculiar result realizes the WGC in this particular setting in a non-trivial fashion: while brane nucleation in non-supersymmetric $\ads$ has been thoroughly investigated~\cite{Maldacena:1998uz,Seiberg:1999xz,Ooguri:2016pdq}, let us stress that in the present case this phenomenon arises from extremal branes interacting in the absence of supersymmetry. 

To conclude this section, let us consider probe-regime interactions between D3-branes, whose corresponding near-horizon throat deviates from $\ads \times \ess$ as we have discussed in Section~\ref{sec:d3-branes}. Indeed, since the EFT parameter $\alpha = 0$ in this case, an $\adsts$ solution cannot exist. For more details on the back-reacted geometry, see~\cite{Angelantonj:2000kh}. Our starting point is now the solution in eqs.~\eqref{eq:d3-branes_metric} and~\eqref{eq:d3-branes_correction}. Let us once again embed the probe world-volume parallel to the $x^\mu$, according to
\begin{eqaed}\label{eq:brane_embedding_d3}
    j \, : \quad x^\mu = \zeta^\mu \, , \qquad u = U(\zeta) \, , \qquad \theta^i = \theta^i_0 \, ,
\end{eqaed}
where the coordinate $u$ is to be interpreted as an energy scale.

\begin{figure}[ht]
	\begin{center}
		\includegraphics[width=.8\linewidth]{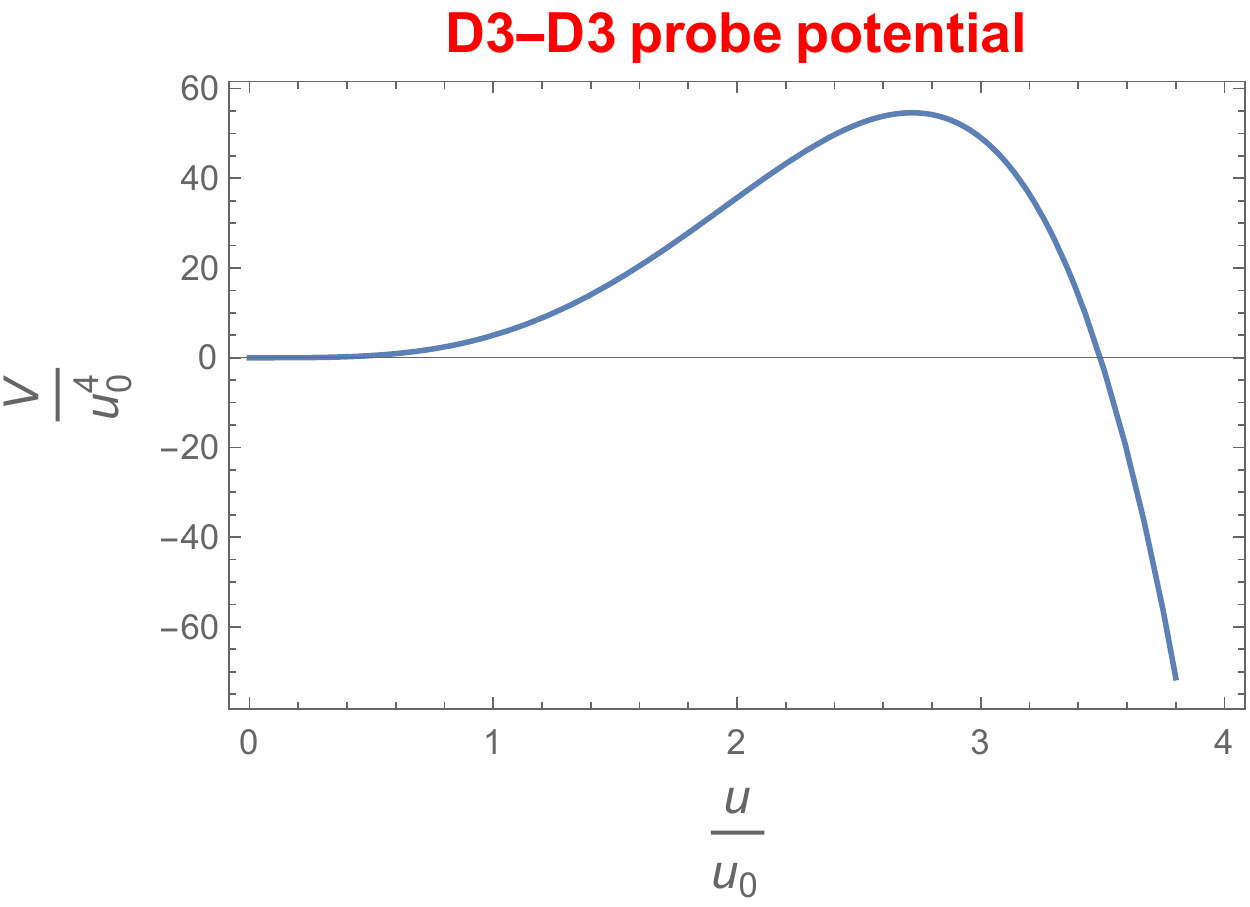}
	\end{center}
	\caption{the normalized probe potential in eq.~\eqref{eq:d3_probe_potential_asymp} in units of the reference scale $u_0$.}
	\label{fig:d3_probe_potential}
\end{figure}

The five-form R-R field strength $F_5$ is self-dual, closed\footnote{Since the orientifold projection removes the Kalb-Ramond form, no additional terms appear in the Bianchi identity.}, and reads
\begin{eqaed}\label{eq:d3_f5}
    F_5 & = \left(1 + \star\right) f_5 \, N \, \text{vol}_{\ess^5} \\
    & = f_5 \, N \, \text{vol}_{\ess^5} + \frac{f_5 \, N}{\widetilde{R}(u)^5} \left( \frac{\alpha' \, u}{R(u)}\right)^3 \, d(\alpha' u) \wedge d^4 x
\end{eqaed}
with $\text{vol}_{\ess^5}$ the volume form of the unit $5$-sphere. The flux quantization condition
\begin{eqaed}\label{eq:f_5_flux_quantization}
    \frac{1}{2\kappa_{10}^2} \int_{\ess^5} F_5 = \mu_3 \, N
\end{eqaed}
then fixes
\begin{eqaed}\label{eq:f5_parameter}
    f_5 = \frac{2\kappa_{10}^2 \, \mu_3}{\Omega_5} \, ,
\end{eqaed}
and the relevant contribution to the potential $C_4$, to be pulled back on the probe world-volume, takes the form
\begin{eqaed}
    C_4 = c_4(u) \, d^4 x + \dots
\end{eqaed}
where $dC_4 = F_5$ implies
\begin{eqaed}
    \frac{c_4'(u)}{\alpha'} = \frac{f_5 \, N}{\widetilde{R}(u)^5} \left( \frac{\alpha' \, u}{R(u)}\right)^3 \, .
\end{eqaed}
Collecting all the ingredients, and using the string-frame world-volume action of eq.~\eqref{eq:world-volume_action}, the probe potential evaluates to
\begin{eqaed}\label{eq:d3_probe_potential}
    V_{\text{probe}}^{\text{D}3}(U) = T_3 \left( \frac{\alpha' \, U}{R(U)} \right)^4 e^{-\phi(U)} - \mu_3 \, c_4(U) \, ,
\end{eqaed}
and its dominant contribution in the EFT limit $g_s \, , g_s^2 \, N \ll 1$, $g_s \, N \gg 1$ is
\begin{eqaed}\label{eq:d3_probe_potential_asymp}
    \frac{V_{\text{probe}}^{\text{D}3}(U)}{U^4} & \sim \frac{16 \, \pi \, \alpha'^2 \, T_3 - f_5 \, \mu_3}{64 \, \pi^2 \, g_s^2 \, N} + \frac{15 \, f_5 \, \mu_3 \, \alpha' T}{8192 \sqrt[4]{8} \, \pi^2} \\
    & + \frac{3 \left(64 \, \pi \, \alpha'^2 T_3 - 5 \, f_5 \, \mu_3 \right) \alpha' T}{2048 \sqrt[4]{8} \, \pi^2} \, \log \left(\frac{U}{u_0}\right) \, .
\end{eqaed}
As expected, substituting the supersymmetric values
\begin{eqaed}\label{eq:d3_susy_values}
    2\kappa_{10}^2 = (2\pi)^7 \, \alpha'^4 \, , \qquad T_3 = \mu_3 = \frac{N_3}{(2\pi)^3 \, \alpha'^2}
\end{eqaed}
for $N_3 \ll N$ probes, and using eq.~\eqref{eq:f5_parameter}, the leading term vanishes, on account of the BPS property, while the remaining sub-leading terms reflect supersymmetry breaking and their $U$-dependence simplifies to
\begin{eqaed}\label{eq:d3_probe_sub-leading}
    V_{\text{sub-leading}}^{\text{D}3}(U) \propto U^4 \left[5 - 4 \, \log \left(\frac{U}{u_0}\right) \right] \, .
\end{eqaed}
Once again \emph{this potential is repulsive}, realizing the WGC also in this ``marginal'' setting, since $U \to \infty$ corresponds to exiting the throat. Interestingly, as depicted in fig.~\ref{fig:d3_probe_potential}, the potential in eq.~\eqref{eq:d3_probe_sub-leading} features a maximum at $U = e \, u_0$ before crossing zero at $U = e^{\frac{5}{4}} \, u_0$, and the height of the potential barrier scales proportionally to $u_0^4$. Therefore, even if the probe stack were initially located in the classically attractive region, it would eventually tunnel to the repulsive region. Indeed, for suitably small velocities $\alpha'\partial_t U \ll 1$, the kinetic term in the DBI action is asymptotic to
\begin{eqaed}\label{eq:d3_kinetic_dbi}
    \frac{\alpha'^2 \, T_3}{2\,g_s} \left[1 + \frac{3 \, \alpha' T}{8 \sqrt[4]{8}} \, g_s^2 \, N \, \log \left(\frac{U}{u_0}\right)\right] \left(\partial_t U\right)^2 \, ,
\end{eqaed}
and since the potential is purely sub-leading the canonical field redefinition would not modify its dominant contribution. Furthermore, let us recall that the geometry in eq.~\eqref{eq:d3-branes_correction} has a limited regime of validity, which we expect to extend at most up to $\frac{u}{u_0} = \mathcal{O}(e^N)$, where sensible dynamics is expected to replace the unbounded approximate potential of eq.~\eqref{eq:d3_probe_potential_asymp}. A potentially instructive toy model, in which the unknown large-$u$ dynamics is substituted by a hard wall at some cut-off scale $u = \Lambda$, would then feature an asymmetric double-well, which can result in bubble nucleation~\cite{Coleman:1977py, Callan:1977pt}.

As we have discussed in the preceding section, an attempt to reproduce these results via a string amplitude computation, at least for the orientifold models, would be met by considerable difficulties, since the relevant annulus contribution vanishes. On the other hand, in the non-extremal case one has access both to the gravitational back-reaction of $\text{D}8$-branes and to a string amplitude computation, and thus we shall turn to this issue in Section~\ref{sec:probe_dm}.

As a final comment, let us observe that in the heterotic model one can also compute the potential for probe $\text{F}1$-strings, extended along one of the directions parallel to the $\text{NS}5$-branes. However, the Kalb-Ramond form $B_2$ vanishes upon pull-back on the string world-sheet, and thus the resulting force is attractive. The counterpart of this setting in the orientifold models would involve probe $\text{D}5$-branes, but these would wrap contractible cycles on the spheres, leading to an uncontrolled computation.

\subsubsection{Brane probes in the Dudas-Mourad geometry}\label{sec:probe_dm}

Let us now extend the considerations of the preceding section to the case in which at least one of the two brane stacks is uncharged. While this case is not directly relevant for the WGC, it is instructive to compare the resulting dynamics to string amplitude computations in the absence of supersymmetry. Indeed, with respect to the extremal case, the leading contribution to the relevant string amplitude corresponds to annulus, which does not vanish and does not entail Riemann surfaces of higher Euler characteristic and other complications. In particular, we shall focus on D$8$-branes in the orientifold models, since their back-reacted geometry is described by the static Dudas-Mourad solution\footnote{The generalization to non-extremal branes of different dimensions would involves non-integrable Toda-like systems, whose correct boundary conditions are not well-understood hitherto. In addition, a reliable probe regime would exclude the pinch-off asymptotic region, thereby requiring numerical computations.}~\cite{Dudas:2000ff} that we have described in Section~\ref{sec:static_solutions}.

Furthermore, the other controlled back-reacted geometry in this setting corresponds to $\text{D}1$-branes, and D$8$-branes are the only other probes whose potential can be reliably computed in this case, since they can wrap the $\ess^7$ in the near-horizon $\ads_3 \times \ess^7$ throat. On the other hand, as we have discussed, while similar considerations apply to the heterotic model a microscopic interpretation appears more subtle. Nevertheless, probe-brane calculations in this setting yield attractive potentials for $8$-branes and fundamental strings, as in the orientifold models, while $\text{NS}5$-branes are repelled. In addition, in some cases the potential scales with a positive power of $g_s$. At any rate, the instability appears to be under control, since probes would reach the strong-coupling regions in a parametrically large time for $g_s \ll 1$.

To begin with, we shall consider a stack of $N_p$ D$p$-branes probing the Dudas-Mourad geometry. According to the considerations in the preceding sections, this ought to describe a stack of D$p$-branes parallel to a stack of D$8$-branes\footnote{By analogy with the results in~\cite{Antonelli:2019nar}, the number $N_8$ of D$8$-branes, should be implicitly determined by the only free parameter $g_s \equiv e^{\Phi_0}$ in eq.~\eqref{eq:dm_orientifold_einstein_mod}, with $g_s \ll 1$ for $N_8 \gg 1$.}. In order to simplify the ensuing discussion, we work in the string frame in units where $\alpha_{\text{O}} = 1$. Since the boundary of the interval spanned by the coordinate $y$ hosts two singularities, we expect this configuration to be under control insofar as the (string-frame) geodesic coordinate
\begin{eqaed}\label{eq:geodesic_r}
	r \equiv \frac{1}{\sqrt{g_s}} \, \int_0^y \frac{du}{u^{\frac{1}{3}}} \, e^{- \, \frac{3}{8} \, u^2}
\end{eqaed}
is far away from its endpoints $r=0$, $r=R_c$. This overlap regime indeed exists, provided that $g_s \equiv e^{\Phi_0} \ll 1$.

\begin{figure}[ht!]
	\begin{center}
		\includegraphics[width=\linewidth]{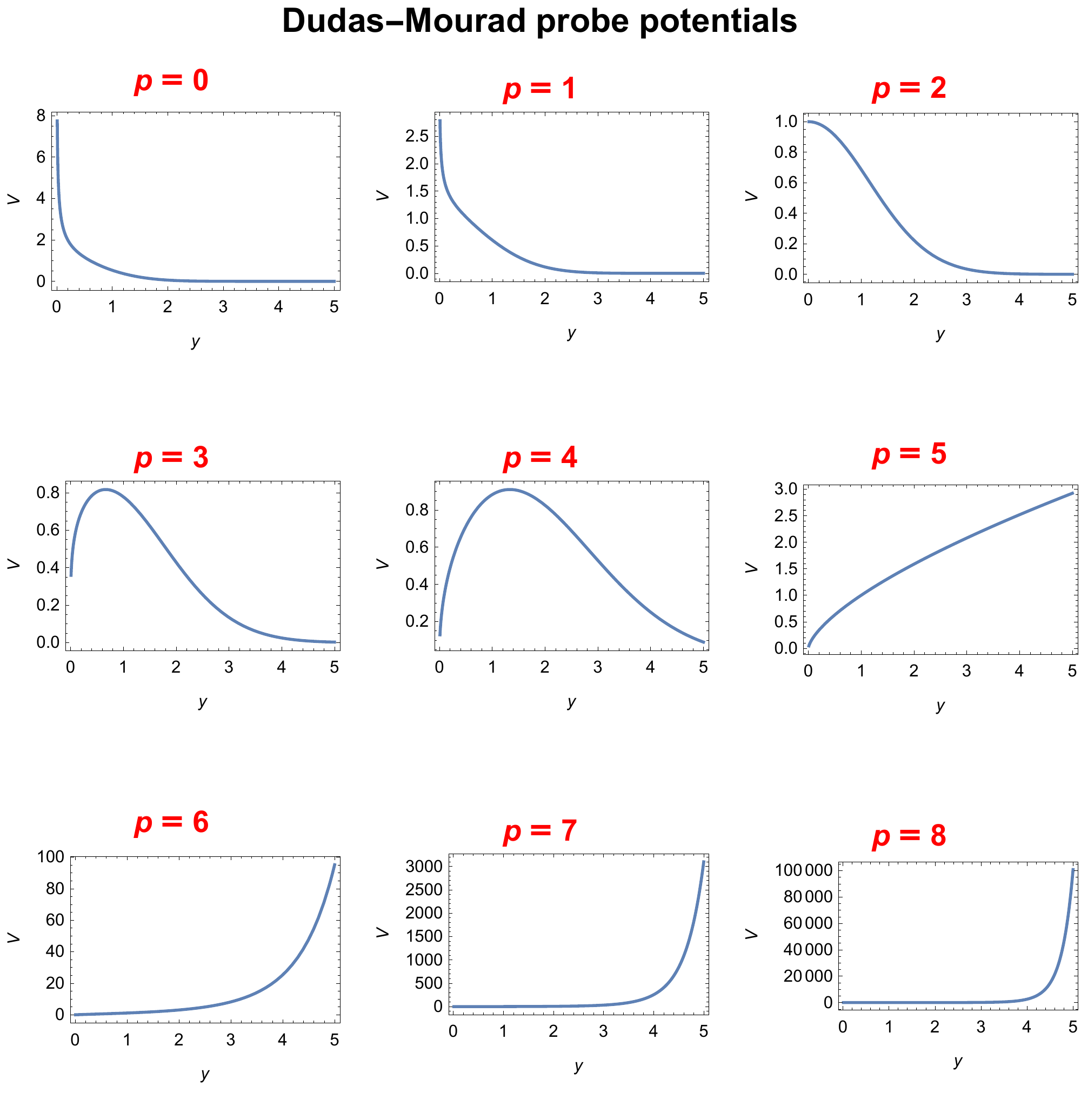}
	\end{center}
	\caption{probe potentials for $g_s = 1$ and $p \leq 8$. For $p < 3$ the probe stack is repelled by the D8-branes, while for $p > 4$ it is attracted to the D$8$-branes. A string amplitude computation yields a qualitatively similar behavior, despite the string-scale breaking of supersymmetry.}
	\label{fig:probe_potential_dm_y}
\end{figure}

\begin{figure}[ht!]
	\begin{center}
		\includegraphics[width=\linewidth]{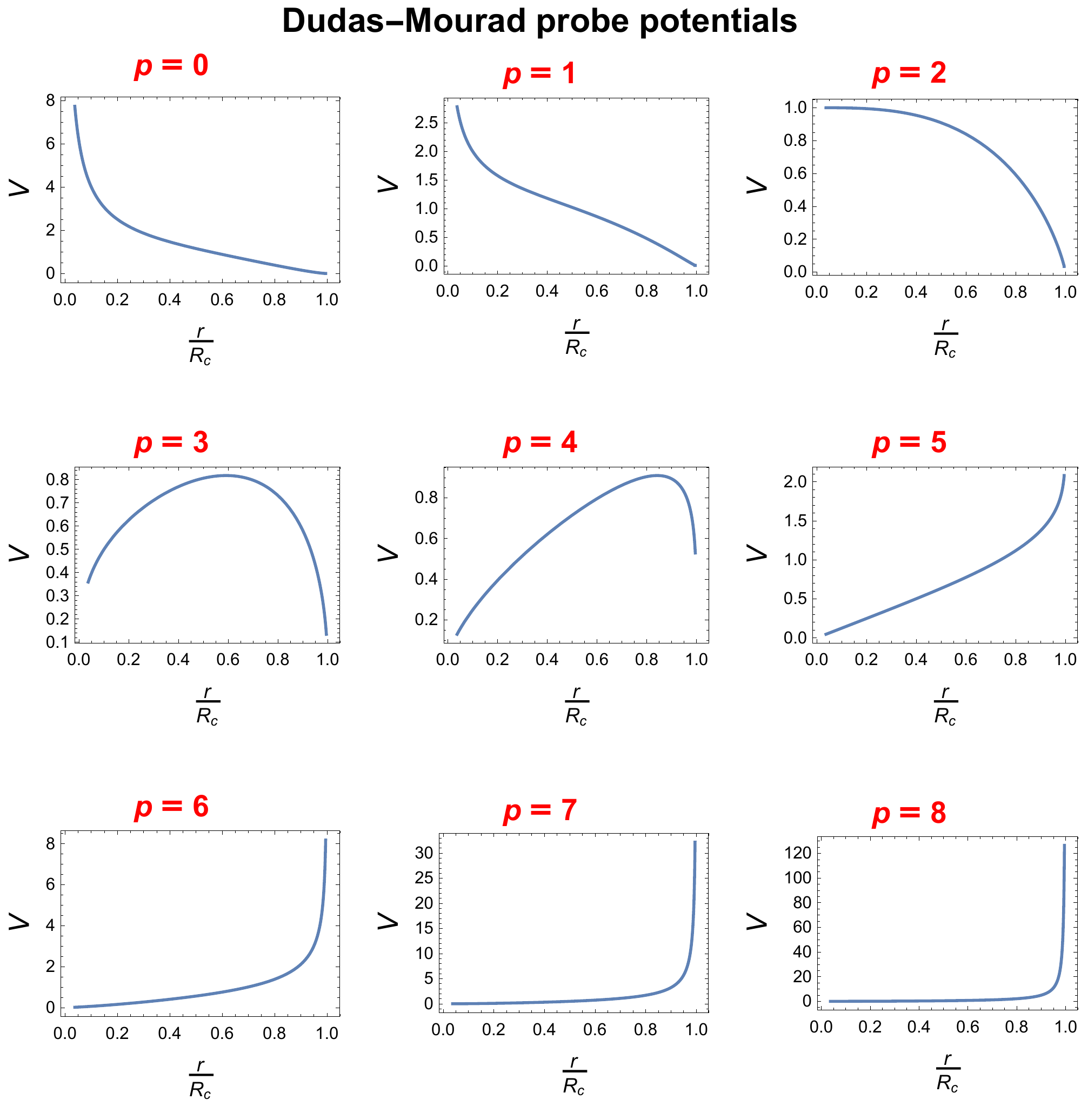}
	\end{center}
	\caption{probe potentials for $g_s = 1$ and $p \leq 8$, plotted as functions of the geodesic coordinate along the compact direction.}
	\label{fig:probe_potential_dm_r}
\end{figure}

Defining the (string-frame) warp factors $A(y) \, , \, B(y)$ of~\cite{Dudas:2000ff} according to
\begin{eqaed}\label{eq:dm_metric_AB}
	ds^2_{10} = e^{2A(y)} \, dx_9^2 + e^{2B(y)} \, dy^2 \, ,
\end{eqaed}
the probe action for D$p$-branes reduces to the DBI action in the absence of fluxes, and it evaluates to
\begin{eqaed}\label{eq:dbi_dm}
	S_p & = - \, N_p \, T_p \int d^{p+1}x \, e^{(p+1)A(y) - \Phi(y)} \\
		& \equiv - \, N_p \, T_p \int d^{p+1}x \, V_{p8} \, ,
\end{eqaed}
where we have defined the potential per unit tension
\begin{eqaed}\label{eq:probe_potential_dm}
	V_{p8} = {g_s}^{\frac{p-3}{4}} \, y^{\frac{2}{9} \left(p-2\right)} \, e^{\frac{p-5}{8} \, y^2} \, .
\end{eqaed}
The non-trivial dependence of eq.~\eqref{eq:probe_potential_dm} on $p$ is depicted in figs.~\ref{fig:probe_potential_dm_y} and~\ref{fig:probe_potential_dm_r}. If the potential drives probes toward $y \; \to \; \infty$ it is repulsive, since the pinch-off this regime hosts the pinch-off singularity discussed in~\cite{Antonelli:2019nar}. As a result, for $p < 3$ probes are repelled by the D$8$-branes, while for $p > 4$ they are attracted to the D$8$-branes. The cases $p = 3 \, , \, 4$ exhibit unstable equilibria, but at large separations the potentials are repulsive. This is the regime that we shall compare with a string amplitude computation.


As we have anticipated, branes probing the back-reacted geometry sourced by non-extremal $p$ branes, with $p < 8$, entail considerable difficulties. Indeed, even if a reliable regime were numerically under control, the correct asymptotic boundary conditions pertaining to branes are yet to be found. In contrast, for extremal branes one can exploit the fact that the near-horizon throat should be deep enough to ensure the reliability of the probe regime.

In order to further investigate the parametric control of our results, one can verify that the probe stack remains far away from the boundary of the interval for parametrically large times. To this end, one can consider rigid branes moving along $y$, starting from the initial conditions $y(0) = y_0 \, , \, \dot{y}(0) = 0$. The reduced Lagrangian
\begin{eqaed}\label{eq:dbi_dm_reduced}
	\mathcal{L}_{\text{red}} = - T_p \, N_p \, V_{p8} \, \sqrt{1 - e^{2\left(B - A \right)} \, \dot{y}^2}
\end{eqaed}
leads to the conserved Hamiltonian
\begin{eqaed}\label{eq:dbi_hamiltonian}
	H_{\text{red}} = \frac{T_p \, N_p \, V_{p8}}{ \sqrt{1 - e^{2\left(B - A \right)} \, \dot{y}^2}} = T_p \, N_p \, V_{p8}(y_0) \, ,
\end{eqaed}
and thus, solving by quadrature, one finds that
\begin{eqaed}\label{eq:time_separated}
	t = \int_{y_0}^y \frac{e^{B(u)-A(u)}}{\sqrt{1 - \left(\frac{V_{p8}(u)}{V_{p8}(y_0)}\right)^2}} \, du = \int_{y_0}^y \frac{g_s^{- \frac{3}{4}} \, e^{- \frac{u^2}{2}}}{u^{\frac{5}{9}} \, \sqrt{1 - \left(\frac{u}{y_0}\right)^{\frac{4}{9}\left(p-2\right)} \, e^{\frac{p-5}{4} \left(u^2 - y_0^2\right)}}}
\end{eqaed}
is indeed parametrically large in string units for $g_s \ll 1$.

\subsubsection{Probe 8-branes in \texorpdfstring{$\ads \times \ess$}{AdS x S} throats}\label{sec:probe_8}

To conclude our analysis of probe branes, let us finally consider $N_8$ D$8$-branes probing the near-horizon geometries sourced by $N_1 \gg N_8$ extremal $\text{D}1$-branes or $N_3 \gg N_8$ extremal $\text{D}3$-branes in the orientifold models. For completeness we shall also consider 8-branes probing the $\ads_7 \times \ess^3$ throat sourced by $N_5 \gg N_8$ $\text{NS}5$-branes in the heterotic model. These comprise the last settings that we shall consider, since the $8$-branes can wrap the internal spheres without collapsing in a vanishing cycle, while leaving enough dimensions to be parallel to the heavy stack. These are also the only cases where computations can be compared to the results in the preceding sections, which hold in the opposite regime $N_1 \, , \, N_3 \, , \, N_5 \ll N_8$. The respective potentials $V_{81} \, , \, V_{83} \, , \, V_{85}$ arise from the DBI contribution only, and take the form
\begin{eqaed}\label{eq:8-brane_probe_potentials}
	V_{81} & \propto N_8 \, T_8 \, R^7 \left( \frac{L}{Z} \right)^2 \, , \\
	V_{85} & \propto N_8 \, T_8 \, R^3 \left( \frac{L}{Z} \right)^6 \\ 
\end{eqaed}
for the $\ads \times \ess$ throats of eq.~\eqref{eq:adsxs_ansatz}, up to an irrelevant (positive) constant, while
\begin{eqaed}\label{eq:8-brane_probe_potential_3}
	V_{83} \sim \sqrt{2} \, \pi^{\frac{1}{4}} \, {\alpha'}^{\frac{9}{2}} \, g_s^{-\frac{3}{4}} \, N_3^{\frac{1}{4}} \, N_8 \, T_8 \, U^4 \left(1 + \frac{3}{8} \, g_s \, T \, \log\left(\frac{U}{u_0}\right) \right)
\end{eqaed}
in the near-horizon limit of the geometry of eq.~\eqref{eq:d3-branes_metric} sourced by D3-branes in the type $0'$B model. These potentials are attractive, which may at first glance appear in contradiction with the results in the following section. However, at large separation, 8-branes wrapped around the internal spheres should behave as (uncharged) $1$-branes, $5$-branes and $3$-branes respectively, consistently with an attractive potential between branes of equal dimension.

\subsection{String-amplitude regime}\label{sec:string_amplitude}

We can now compare the results of the preceding probe-brane analysis for uncharged branes to a string amplitude computation. The relevant leading-order amplitude encoding the interaction between parallel stack of $N_p$ $\text{D}p$-branes and $N_q$ $\text{D}q$-branes\footnote{One can expect that an amplitude computation be reliable for $N_p \, , N_q = \mathcal{O}\!\left(1\right)$, in contrast to the probe regimes $N_p \gg N_q$ and $N_p \ll N_q$.}, with $p < q$ for definiteness, corresponds to the annulus. The transverse-channel integrand in the present cases takes the form~\cite{Dudas:2001wd}
\begin{eqaed}\label{eq:A_pq}
	\widetilde{\mathcal{A}}_{pq} \propto N_p \, N_q \left( V_{8-q+p} \, O_{q-p} - O_{8-q+p} \, V_{q-p} \right)
\end{eqaed}
up to a (positive) normalization, where the characters are evaluated at $\mathfrak{q} = e^{-2\pi \ell}$. In suitable units for the transverse separation $r$, the potential $V_{pq}$ is then given by
\begin{eqaed}\label{eq:string_potential}
	V_{pq} \propto - \, N_p \, N_q \int_0^\infty \frac{d\ell}{\ell^{\frac{9-q}{2}}} \, \frac{\widetilde{A}_{pq}}{\eta^{8-q+p}} \left( \frac{2\eta}{\vartheta_2} \right)^{\frac{q-p}{2}} e^{- \, \frac{r^2}{\ell}} \, .
\end{eqaed}
For large $r$, the integral is dominated by the large-$\ell$ region. In this region eq.~\eqref{eq:A_pq} asymptotes to $\mathfrak{q}^{-\frac{1}{3}} \, \tilde{A}_{pq}$, with
\begin{eqaed}\label{eq:amplitude_integrand_asymptotics}
	\tilde{A}_{pq} & \propto V_{8-q+p} \, O_{q-p} - O_{8-q+p} \, V_{q-p} \\
	& \sim 2 \left( 4 - q + p \right) \mathfrak{q}^{\frac{1}{3}} \, ,
\end{eqaed}
so that for $q < 7$ one finds
\begin{eqaed}\label{eq:large_r_string_potential}
	V_{pq} \propto \left(q - p - 4 \right) \frac{N_p \, N_q}{r^{7-q}} \, .
\end{eqaed}
This potential is repulsive for $p < q - 4$ and attractive for $p > q - 4$. While the integral in eq.~\eqref{eq:string_potential} diverges for $q \geq 7$, a distributional computation for $q = 7 \, , \, 8$ yields a finite force stemming from potentials proportional to $\left(p - 3 \right) \log(r)$ and $\left(p - 4 \right) r$ respectively. In order to compare these results to the probe-brane analysis of the preceding section, the relevant cases are thus $p=q$, which leads to an attractive potential, consistently with eqs.~\eqref{eq:8-brane_probe_potentials} and~\eqref{eq:8-brane_probe_potential_3}, and $q = 8$, which leads to a potential proportional to $(p - 4) \, r$. Therefore, the latter interaction is repulsive for $p < 4$ and attractive for $p > 4$, consistently with the results in the preceding section. Let us remark that, in the absence of (linear) supersymmetry, as well as a flat space-time background, the agreement between the two computations, performed in complementary regimes, seems quite non-trivial, and suggests a deeper principle akin to the WGC, or the repulsive force conjecture, for uncharged extended objects.

\subsection{Holographic regime}\label{sec:world-volume_gauge}

The appearance of (unstable) $\ads$ geometries in the models that we have investigated, together with their connection to non-supersymmetric brane configurations, suggests that the world-volume gauge theories living on the branes could encode their dynamics holographically, including gravitational instabilities. Indeed, the perturbative instabilities studied in~\cite{Gubser:2001zr,Basile:2018irz} ought to correspond to operators with complex anomalous dimension~\cite{Klebanov:1999um, DeWolfe:2001nz}, while the holographic description of non-perturbative instabilities is more subtle~\cite{deHaro:2006ymc, Papadimitriou:2007sj, Maldacena:2010un, Barbon:2010gn, Barbon:2011zz, Harlow:2010az, Maxfield:2014wea, Bhattacharyya:2017pqq, Burda:2018rpb, Hirano:2018cyr} and has been proposed to be captured by dual RG flows~\cite{Antonelli:2018qwz, Ghosh:2021lua}. In the latter picture, the putative $\cft$ deformations ought to be ``heavy'', since their effect is suppressed in the large-flux limit, but some properties of brane dynamics can be understood holographically, if qualitatively, in the weak-coupling regime of the gauge theory. 

Concretely, the weak-gravity repulsive effect that we have discussed in Section~\ref{sec:probe_potentials} suggests that the gauge group undergoes a dynamical breaking according to~\cite{Witten:1998xy,Seiberg:1999xz}
\begin{eqaed}\label{eq:gauge_group_reduction}
	U(N) \; & \to \; U(N - \delta N) \times U(\delta N) \, , \\
	USp(2N) \; & \to \; USp(2N - 2\delta N) \times USp(2\delta N)
\end{eqaed}
in the type $0'$B model and the Sugimoto model respectively. However, the initial expectation value attained by scalars upon separating the branes would increase due to the repulsive force, and thus a conventional Higgs mechanism would not appear to be involved, at least not in its usual guise. While the bulk $\ads$ geometry is expected to be dual to the IR regime of the world-volume gauge theory, which as we shall see is strongly coupled, one can make progress studying its weakly coupled UV counterpart, under the assumption that the qualitative character of brane interactions be preserved under RG flow. In the Sugimoto model, the relevant gauge theory arises projecting a supersymmetric one. In settings of this type the projected theory retains some properties of the parent theory~\cite{Kachru:1998ys, Lawrence:1998ja, Bershadsky:1998mb, Bershadsky:1998cb, Schmaltz:1998bg, Erlich:1998gb, Silverstein:2000ns, Tong:2002vp}, and we intend to explore this intriguing idea in future work. However, for the time being we shall focus on the weakly coupled UV regime of this gauge theory, alongside its $U(N)$ counterpart in the type $0'$B model\footnote{For similar considerations on D3-branes in the type $0'$B model, for which the gauge theory is strongly coupled in the UV, see~\cite{Dudas:2000sn}.}.

To begin with, one is to address the issue of which background the branes are placed in. According to general considerations along the lines of~\cite{Maldacena:1997re}, one ought to place the branes in the flux-less limit of the back-reacted geometry of eq.~\eqref{eq:adsxs_ansatz}. However, in the absence of supersymmetry the resulting configuration appears highly curved and far outside the EFT regime of validity, and in particular there is no Minkowski solution to replace it. However, introducing $N_8 \gg 1$ D$8$-branes sourcing the static Dudas-Mourad geometry of eq.~\eqref{eq:dm_orientifold_einstein_mod}, the situation appears more tame, since $N \gg N_8$ $\text{D}1$-branes placed in the controlled region described in Section~\ref{sec:probe_dm} would dominate the back-reaction. If this construction is reliable, standard decoupling arguments~\cite{Maldacena:1997re} should lead to a two-dimensional world-volume gauge theory on flat space-time. In the case of the Sugimoto model, the massless perturbative spectrum has been described in~\cite{Sugimoto:1999tx}, while the case of the type $0'$B model was studied in~\cite{Dudas:2000sn}. 

The corresponding world-volume (Euclidean) effective action $S_{\text{D}1}$ then takes the schematic form
\begin{eqaed}\label{eq:d1_eff_action}
	S^E_{\text{D}1} = \frac{1}{g_\text{YM}^2} \, \text{Tr} & \int d^2 \zeta \, \bigg( \left(\partial_+ A_-\right)^2 + \partial_+ X_i \left[\mathcal{D}_- \, , X_i \right] - \, \frac{1}{4} \left[X_i \, , X_j \right] \left[X_i \, , X_j \right] \\
	& + \psi_+ \left[\mathcal{D}_- \, , \psi_+ \right] + \psi_- \, \partial_+ \psi_- + \psi_- \, \Gamma_i \left[X_i \, , \psi_+ \right] + \lambda_-^A \, \partial_+ \, \lambda_-^A\bigg)
\end{eqaed}
in the (Euclidean) light-cone gauge $A_+ = 0$. In the Sugimoto model, in contrast to its supersymmetric counterpart, the scalars $X_i$ which comprise a vector of the transverse rotation group $SO(8)$ are in the anti-symmetric representation of $USp(2N)$, while the adjoint is symmetric and the world-volume fermion $\psi_+$ (resp. $\psi_-$) is in the symmetric (resp. anti-symmetric) representation and is a $SO(8)$ spinor. The $\lambda_-^A$ are bifundamental fermions of $USp(2N) \times USp(2N_f)$ with $N_f = 16$ ``flavors'', and arise from (massless modes of) open strings stretching between the $\text{D}1$-branes and the $\overline{\text{D}9}$-branes. The type $0'$B model is analogous \textit{mutatis mutandis}, with the important difference that the scalars are in the adjoint representation of the gauge group $U(N)$, since the Möbius strip contribution does not modify the structure encoded in the annulus due to the vanishing O9-plane tension. While the light-cone gauge is convenient, since in two dimensions ghosts decouple~\cite{tHooft:1974pnl} and the gauge field can be integrated out exactly\footnote{Non-perturbative methods of this type in light-cone gauge have been recently employed in~\cite{Dempsey:2021xpf}.}, here we shall content ourselves with a one-loop analysis. In this respect, the $\beta$ function of the gauge coupling depends only on the (perturbative) matter content. In order to derive it, let us recall that, in four dimensions, the result
\begin{eqaed}\label{eq:beta_4d}
	\beta_{4d} = b_1 \, \frac{g_\text{YM}^3}{16 \pi^2}
\end{eqaed}
arises from the (dimension-independent) $a_4$ coefficient in the heat-kernel expansion of the one-loop functional determinant~\cite{Vassilevich:2003xt}. Therefore, in the two-dimensional case the bare coupling $g_0$ would be related to the renormalized coupling according to
\begin{eqaed}\label{eq:bare_coupling_2d}
	\frac{1}{g_0^2} = \frac{1}{g_\text{YM}^2(\mu)} - \frac{b_1}{4 \pi} \, \frac{1}{\mu^2} \, .
\end{eqaed}
In terms of the dimensionless coupling $g_\text{YM} \equiv \widehat{g} \, \mu$, the one-loop $\beta$ function is then
\begin{eqaed}\label{eq:beta_2d}
	\widehat{\beta}_{2d} = - \, \widehat{g} + \frac{b_1}{4 \pi} \, \widehat{g}^3 \, ,
\end{eqaed}
with~\cite{Gross:1973id, PhysRevD.9.2259, Yamatsu:2015npn}
\begin{eqaed}\label{eq:b1_symplectic}
	b_1^{USp(2N)} = \frac{9 \, N +  N_f - 15}{3} \, , \qquad b_1^{U(N)} = \frac{9 \, N + 2 \, N_f}{3}
\end{eqaed}
for the Sugimoto model and the type $0'$B model respectively. Therefore, the gauge coupling eventually flows to a strongly coupled region, which could exhibit confinement or screening~\cite{Frishman:1997uu}, even in the large-$N$ limit where the suitable parameter is the 't Hooft coupling. From the perspective of the bulk, the IR behavior of the gauge coupling ought to reflect the radial perturbations of the dilaton described in~\cite{Antonelli:2019nar}. Specifically, power-like perturbations $\phi - \phi_{\ads} \propto \abs{r}^{- \lambda}$ about the fixed-point $\ads$ throat, encoded in eqs.~\eqref{eq:brane_full_ansatz} and~\eqref{eq:ads_s_solution}, would translate into the IR $\beta$ function
\begin{eqaed}\label{eq:beta_ir}
	\widehat{\beta}_\text{IR} = \left(2 \, \lambda - 1 \right) \left(\widehat{g} - \widehat{g}^*\right) \, ,
\end{eqaed}
since $r$ is related to the Poincar\'e-patch holographic coordinate $z \mapsto \frac{1}{\mu}$ according to $\abs{r} = \frac{z^{2}}{2}$. In particular, the two independent power-like perturbations that decay in the IR yield~\cite{Antonelli:2019nar}
\begin{eqaed}\label{eq:beta_ir_orientifolds}
	\widehat{\beta}_\text{IR} = \sqrt{5} \left(\widehat{g} - \widehat{g}^*\right) \qquad \text{or} \qquad \widehat{\beta}_\text{IR} = \sqrt{13} \left(\widehat{g} - \widehat{g}^*\right)
\end{eqaed}
in the orientifold models. The overall picture appears daunting, but the (non-local) quartic effective action obtained integrating out the gauge field is potentially amenable to large-$N$ Hubbard-Stratonovich techniques~\cite{Weinberg:1997rv, Moshe:2003xn} or non-Abelian bosonization~\cite{Witten:1983ar}.

As we have anticipated, one can still attempt to make some progress studying the UV regime computing the one-loop effective potential. To this end, in order to connect with a space-time interpretation, we shall consider ``geometric'' configurations that describe two parallel stacks indexed by $\alpha = 1 \, , \, 2$ and transverse positions $\mathbf{x}^{(\alpha)}$. The corresponding configuration for the scalars $X_k$ is
\begin{eqaed}\label{eq:brane_separation}
    X_k = \frac{g_{\text{YM}}}{\sqrt{2N}} \, \sum_{\alpha = 1}^2 \Omega_\alpha \, x_k^{(\alpha)} \, ,
\end{eqaed}
where the $\Omega_\alpha$ are block-diagonal. In detail, they are the projections onto the first and second stacks of the (symplectic-)trace singlet, which is the symplectic matrix $i\Omega$ in the Sugimoto model and the identity matrix in the type $0'$B model. The resulting quadratic kinetic operator yields a functional determinant akin to that studied in~\cite{Zarembo:1999hn, Tseytlin:1999ii}, and using a covariant, Feynman-like gauge the contributions of the gauge field cancels that of the ghosts. This is to be expected on account of our preceding considerations. Furthermore, the fermionic terms also vanish in the Sugimoto model, while in the type $0'$B they cancel the bosonic contribution, mirroring the results of~\cite{Tseytlin:1999ii} for self-dual $\text{D}3$-branes. All in all, for the Sugimoto model the quadratic action for fluctuations $\delta X_i \equiv \delta X_i^a \, t_a$, decomposed in an orthogonal basis $\{t_a\}$ of the space of (imaginary) anti-symmetric matrices, takes the form
\begin{eqaed}\label{eq:quadratic_d1}
	S_{\text{D}1}^{(2)} = \int d^2 \zeta \left( \partial_+ \delta X_i^a \, \partial_- \delta X_i^a + \frac{1}{2} \, \delta X_i^a \left(M^2\right)_{ij}^{ab} \delta X_j^b\right) \, ,
\end{eqaed}
where the (positive semidefinite) mass matrix
\begin{eqaed}\label{eq:d1_mass_matrix}
	\left(M^2\right)_{ij}^{ab} = \frac{g_\text{YM}^2}{N} \, \delta_{ij} \, \sum_{\alpha \, , \, \beta} \mathbf{x}^{(\alpha)} \cdot \mathbf{x}^{(\beta)}  \, \omega^{(\alpha \beta)}_{ab} \, , \qquad \omega^{(\alpha \beta)}_{ab} \equiv \text{Tr}\left( \left[\Omega_\alpha \, , t_a\right] \left[\Omega_\beta \, , t_b\right] \right)
\end{eqaed}
arises from the quartic potential of eq.~\eqref{eq:d1_eff_action}, and one is thus led to the one-loop effective potential
\begin{eqaed}\label{eq:1-loop_pot_d1}
	V_{\text{D}1}^{(1)} = \frac{1}{2} \, \text{Tr} \int \frac{d^2p}{\left( 2 \pi\right)^2} \, \log\! \left(p^2 + M^2 \right) \, .
\end{eqaed}
The non-vanishing eigenvalues of the mass matrix comprise not only the separation $r^2 \equiv \abs{\mathbf{x}^{(1)} - \mathbf{x}^{(2)}}^2$ between the stacks, as in the case of orthogonal and unitary gauge groups, but also (twice) the mean position $\abs{\mathbf{x}^{(1)} + \mathbf{x}^{(2)}}^2$. Furthermore, when a stack contains more than one brane, there are eigenvalues proportional to the square $\abs{\mathbf{x}^{(\alpha)}}^2$ of its position. While we were not able to provide a complete explanation of this behaviour, which is due to the non-Abelian nature of the (projected) symplectic matrices $\Omega_\alpha$ and thus reminiscent of brane polarization~\cite{Myers:1999ps}, when interpreted in terms of smooth space-time geometry the apparent breaking of translational symmetry could presumably be ascribed to the absence of a flat background in which the branes can be placed, along with the presence of a force acting on them. This resonates with our preceding considerations, whereby the world-volume gauge theory can be constructed placing probe branes in the controlled region of the Dudas-Mourad geometry: the force acting on probes, which we have discussed in Section~\ref{sec:probe_dm}, appears to be encoded in the one-loop effective potential of the world-volume gauge theory, at least for the Sugimoto model.

Focusing on the contribution to the trace in eq.~\eqref{eq:1-loop_pot_d1} that depends on the brane separation $r$, one finds
\begin{eqaed}\label{eq:1-loop_pot_series}
		V_{\text{D}1}^{(1)}\big|_{\text{sep}} = \frac{m}{\pi} \int_{0}^{\Lambda^2_\text{UV}} ds \, \log\left(s + \frac{g_\text{YM}^2 r^2}{N} \right)
\end{eqaed}
with $m$ the corresponding multiplicity. Notice that the renormalized dimensionless coupling $\widehat{g}(\mu_\text{sep})$ evaluated at the separation scale $\mu_\text{sep}^{-1} \equiv r/\Lambda_\text{UV}$ appears, and the perturbative regime translates into the requirement $\widehat{g}(\mu_\text{sep})\ll 1$, so that
\begin{eqaed}\label{eq:1-loop_asymp}
	V_{\text{D}1}^{(1)}\big|_{\text{sep}} \sim - \, \frac{m}{\pi} \, \frac{g_\text{YM}^2 r^2}{N} \log\!\left( \frac{g_\text{YM}^2 r^2}{N \Lambda_\text{UV}^2} \right)
\end{eqaed}
indeed exhibits a repulsive behavior induced by tunneling. This result, although marginally reliable at the tunneling scale, appears in agreement with axial-gauge computations, and is consistent with our preceding considerations. Moreover, as expected, for $N = 1$ the (gauge singlet) scalars decouple, and thus their effective potential receives no corrections even beyond the one-loop level. This is reflected in eq.~\eqref{eq:1-loop_asymp} by the fact that $m=0$ for $N=1$.
    
\section{Conclusions}\label{sec:conclusions}

    In this paper we have investigated the interactions of branes in non-supersymmetric string models. In particular, we have focused on the $USp(32)$~\cite{Sugimoto:1999tx} and $U(32)$ ``type $0'$B''~\cite{Sagnotti:1995ga, Sagnotti:1996qj} orientifold models and on the $SO(16) \times SO(16)$ heterotic model of~\cite{AlvarezGaume:1986jb, Dixon:1986iz}, computing static interaction potentials for parallel stack of branes. Despite the absence of supersymmetry, the presence of R-R charges entails the existence of charged extremal branes in the spectrum, which comprise $\text{D}1$ and $\text{D}5$-branes in the Sugimoto model and $\text{D}1$, $\text{D}3$, $\text{D}5$ and $\text{D}7$-branes in the type $0'$B model. In addition, we have included heterotic $\text{NS}5$-branes in our analysis, since their behaviour mirrors that of $\text{D}1$-branes to a certain extent. On the other hand, we have excluded $\text{D}5-\text{D}5$ and $\text{D}7-\text{D}7$ interactions, since the corresponding background geometries and string amplitudes present a number of subtleties. Our computations span a variety of regimes: whenever one stack is parametrically heavier than the other, one can replace the heavy stack with the corresponding background geometry probed by the light stack, while whenever both stacks are parametrically light one can compute the interaction potential via the annulus amplitude, provided that the two stacks do not share the same charges. Finally, we have investigated the world-volume gauge theory of $\text{D}1$-branes in the Sugimoto model in an attempt to extend our results beyond the perturbative regime, connecting them to a top-down non-supersymmetric holography of the type discussed in~\cite{Antonelli:2018qwz, Antonelli:2019nar}. We have computed the one-loop $\beta$ function and effective potential for separated branes, finding a departure from the supersymmetric case and from the type $0'$B model.
    
    The interaction potentials that we have obtained are qualitatively consistent among these complementary regimes whenever comparisons are possible. Namely, separated stacks either attract or repel in each case. The agreement is non-trivial, since there is no apparent principle that protects this behavior upon increase of the strength of gravitational backreaction. Moreover, for branes that share the same charge(s), we found a novel mechanism that realizes the Weak Gravity Conjecture via a renormalization of the effective charge-to-tension ratio. This peculiar effect arises from an interaction between the branes mediated by the supersymmetry-breaking ingredients, which at the level of the ten-dimensional EFT are reflected by the gravitational tadpole potential in eq.~\eqref{eq:action}. This behavior is also consistent with the world-volume computations that we have presented in Section~\ref{sec:world-volume_gauge}, although its reliability is confined to the UV due to asymptotic freedom. On the other hand, a proper holographic comparison would entail flowing to the strongly coupled IR, where arguments from the bulk suggest that a large-$N$ fixed point ought to exist. We would like to pursue this direction further in future work, possibly relying on non-perturbative large-$N$ and two-dimensional methods.
    
    All in all, our findings suggest that the instructive lessons that have been gathered from brane dynamics can be robust, to some extent, with respect to the dramatic effects of high-energy supersymmetry breaking. One can expect that deeper connections with holography and Swampland proposals can be unveiled from this perspective, which would nicely complement the bottom-up considerations of~\cite{Lanza:2020qmt, Lanza:2021qsu, Buratti:2021yia}. In this paper we have only taken a first step toward a deeper understanding of these fundamental issues, and a more detailed investigation could encompass $\text{D}5-\text{D}5$ and $\text{D}7-\text{D}7$ interactions, either in the probe regime or in the string-amplitude regime, and more generally probe-brane potentials for the (so far elusive) background geometries sourced by $\text{D}5$ and $\text{D}7$-branes. Other avenues for future research include exploring further the back-reaction and thermodynamics of non-extremal branes, as well as involving higher-order Wess-Zumino terms along the lines of~\cite{Billo:1999bg} to take into account the effects of R-R interactions for branes of different dimensions. Aside from cases where the leading contribution vanishes, one can expect that couplings of this type be required to all orders, at least in $\alpha'$, in order to circumvent a Dine-Seiberg-like argument~\cite{Dine:1985he}. Some progress along these lines has been achieved via all-order string-amplitude computations (see, \emph{e.g.},~\cite{Antonelli:2019pzg, Hatefi:2021czu}) and via T-duality in the context of cosmology~\cite{Meissner:1991zj, Meissner:1996sa, Hohm:2015doa, Hohm:2019ccp, Hohm:2019jgu, Krishnan:2019mkv,Basile:2021amb,Basile:2021krk,Bernardo:2021xtr,Quintin:2021eup}. Furthermore, from the perspective of holography, it would be interesting to understand whether the IR fixed-point structure exhibited by the near-horizon $\ads$ throats discussed in Section~\ref{sec:geometries} can be reproduced via strong-coupling effects in the putative dual gauge theory. While the limit of a large number of branes appears necessary in order to probe this regime, small-$N$ computations seem to reveal intriguing hints concerning the scenarios proposed in~\cite{Antonelli:2018qwz, Antonelli:2019nar} and their connections to the conjectures put forth in~\cite{Ooguri:2006in, Ooguri:2018wrx, Lust:2019zwm, Lee:2019xtm, Lee:2019wij, Baume:2020dqd, Perlmutter:2020buo}. These and other related issues are currently under investigation.

\section*{Acknowledgements}

It is a pleasure to thank C. Angelantonj, E. Dudas, M. Bianchi and A. Sagnotti for helpful feedback during the development of this work. I am also grateful to G. Bogna, S. Raucci, S. Bottaro, D. Bufalini, S. Lanza and A. Faraggi for insightful discussions and feedback on the manuscript.

This work was supported by the Fonds de la Recherche Scientifique - FNRS under Grants No.\ F.4503.20 (``HighSpinSymm'') and T.0022.19 (``Fundamental issues in extended gravitational theories'').


\printbibliography

\end{document}

%% file: geometry2.pdf_tex
\begingroup%
  \makeatletter%
  \providecommand\color[2][]{%
    \errmessage{(Inkscape) Color is used for the text in Inkscape, but the package 'color.sty' is not loaded}%
    \renewcommand\color[2][]{}%
  }%
  \providecommand\transparent[1]{%
    \errmessage{(Inkscape) Transparency is used (non-zero) for the text in Inkscape, but the package 'transparent.sty' is not loaded}%
    \renewcommand\transparent[1]{}%
  }%
  \providecommand\rotatebox[2]{#2}%
  \newcommand*\fsize{\dimexpr\f@size pt\relax}%
  \newcommand*\lineheight[1]{\fontsize{\fsize}{#1\fsize}\selectfont}%
  \ifx\svgwidth\undefined%
    \setlength{\unitlength}{361.85638668bp}%
    \ifx\svgscale\undefined%
      \relax%
    \else%
      \setlength{\unitlength}{\unitlength * \real{\svgscale}}%
    \fi%
  \else%
    \setlength{\unitlength}{\svgwidth}%
  \fi%
  \global\let\svgwidth\undefined%
  \global\let\svgscale\undefined%
  \makeatother%
  \begin{picture}(1,0.51711375)%
    \lineheight{1}%
    \setlength\tabcolsep{0pt}%
    \put(0,0){\includegraphics[width=\unitlength,page=1]{geometry2.pdf}}%
    \put(0.77294008,0.26901672){\color[rgb]{0,0,0}\makebox(0,0)[lt]{\begin{minipage}{0.25538106\unitlength}\centering $\phi\rightarrow \infty$\end{minipage}}}%
    \put(0.37304658,0.48571341){\color[rgb]{0,0,0}\makebox(0,0)[lt]{\begin{minipage}{0.1967578\unitlength}\centering $\int \sqrt{g_{rr}} dr$\end{minipage}}}%
    \put(-0.02475508,0.27922251){\color[rgb]{0,0,0}\makebox(0,0)[lt]{\begin{minipage}{0.25538106\unitlength}\centering $\ads_{p+2}\times\ess^q$\end{minipage}}}%
    \put(-0.00658487,0.24092555){\color[rgb]{0,0,0}\makebox(0,0)[lt]{\begin{minipage}{0.25538106\unitlength}\centering $\phi=\phi_0$\end{minipage}}}%
    \put(0.37029824,0.05353275){\color[rgb]{0,0,0}\makebox(0,0)[lt]{\begin{minipage}{0.11829335\unitlength}\centering $\ess^q$\end{minipage}}}%
    \put(0,0){\includegraphics[width=\unitlength,page=2]{geometry2.pdf}}%
  \end{picture}%
\endgroup%